\documentclass[prd,
			altaffilletter,
			twocolumn,
			superscriptaddress,
			floatfix,
			nofootinbib]{revtex4} 
\usepackage{amsfonts,
			amsmath,
			graphicx,
			bm,
			tikz-feynman,
			float,
			mathtools,
			color,
			bbold,
			url,
			todonotes,
			ifpdf,
			multirow,
			footmisc}
			
\usetikzlibrary{shapes.misc}
\tikzset{cross/.style={cross out, draw=black, minimum size=2*(#1-\pgflinewidth), inner sep=0pt, outer sep=0pt},
cross/.default={3pt}}

\usepackage[colorlinks=true,
			linkcolor=red,
			urlcolor=red,
			citecolor=red]{hyperref}
			
\usepackage[yyyymmdd,hhmmss]{datetime}

\def\today{\number\day\space\ifcase\month\or
January\or February\or March\or April\or May\or June\or
July\or August\or September\or October\or November\or December\fi
\space\number\year}
\newcount\mins \newcount\hours
\def\now{\hours=\time \mins=\time
	\divide\hours by60 \multiply\hours by60 \advance\mins by-\hours
	\divide\hours by60 
	\number\hours:\ifnum\mins<10 0\fi\number\mins }

\newcommand{\cambridge}{Department of Applied Mathematics and Theoretical Physics, University of Cambridge, Cambridge, CB3 0WA, UK}
\newcommand{\glasgow}{SUPA, School of Physics and Astronomy, University of Glasgow, Glasgow, G12 8QQ, UK}
\newcommand{\tehran}{Department of Physics, University of Tehran, Tehran 1439955961, Iran}

\begin{document}
\title{$B_c \to B_{s(d)}$ form factors from lattice QCD}

\author{Laurence J. \surname{Cooper}} 
\email[]{L.J.Cooper@damtp.cam.ac.uk}
\affiliation{\cambridge}
\affiliation{\glasgow}

\author{Christine~T.~H.~\surname{Davies}} 
\affiliation{\glasgow}

\author{Judd \surname{Harrison}} 
\email[]{Judd.Harrison@glasgow.ac.uk}
\affiliation{\glasgow}

\author{Javad \surname{Komijani}} 
\affiliation{\glasgow}
\affiliation{\tehran}

\author{Matthew \surname{Wingate}}
\affiliation{\cambridge}

\collaboration{HPQCD Collaboration}
\email[URL: ]{http://www.physics.gla.ac.uk/HPQCD}

\begin{abstract}
\noindent We present results of the first lattice QCD calculations of $B_c \to B_s$ and $B_c \to B_d$ weak matrix elements. Form factors across the entire physical $q^2$ range are then extracted and 
extrapolated to the physical-continuum limit before combining with CKM matrix 
elements to predict the semileptonic decay rates $\Gamma (B_c^+ \to B_s^0 \overline{\ell} \nu_{\ell})=26.2(1.2) \times 10^{9} \hspace{1mm}\text{s}^{-1}$ and $\Gamma (B_c^+ \to B^0 \overline{\ell} \nu_{\ell})=1.65(10)\times 10^{9}\hspace{1mm} \text{s}^{-1}$. The lattice QCD uncertainty is comparable to the CKM uncertainty here.
Results are derived from correlation functions computed on MILC Collaboration gauge configurations with a range of lattice spacings including 2+1+1 flavours of dynamical sea quarks in the Highly Improved Staggered Quark (HISQ) formalism. HISQ is also used for the propagators of the valence light, strange, and charm quarks. Two different formalisms are employed for the bottom quark: non-relativistic QCD (NRQCD) and heavy-HISQ. Checking agreement between these two approaches is an important test of our strategies for heavy quarks on the lattice. From chained fits of NRQCD and heavy-HISQ data, we obtain the differential decay rates $d\Gamma/ d q^2$ as well as integrated values for comparison to future experimental results.
\end{abstract}

\maketitle


\section{Introduction}

The semileptonic weak decays $B_c^+ \to B_s^0 \overline{\ell} \nu_{\ell}$ and $B_c^+ \to B^0 \overline{\ell} \nu_{\ell}$ proceed via tree-level flavour changing processes $c \to sW^+$ and $c \to dW^+$ parametrised by the Cabbibo-Kobayashi-Maskawa (CKM) matrix of the Standard Model.
Associated weak matrix elements can be expressed in terms of form factors which capture the non-perturbative QCD physics.
Precise determination of the normalisation and the $q^2$ dependence of these form factors from lattice QCD will allow a novel comparison with future experiment to deduce the CKM parameters $V_{cs}$ and $V_{cd}$.
Lattice studies of other semileptonic meson decays that involve tree-level weak decays of a constituent charm quark include~\cite{Aubin:2004ej,Na:2010uf,Koponen:2013tua,Donald:2013pea,Bazavov:2014wgs,Lubicz:2017syv}.
Precise determination of these CKM matrix elements is critical for examining the second row unitary constraint
\begin{align}
|V_{cd}|^2 + |V_{cs}|^2 + |V_{cb}|^2 = 1.
\end{align}
This will complement other unitarity tests of the CKM matrix.
It is possible LHCb could measure $B_c^+ \to B_s^0 \overline{\mu} \nu_{\mu}$ using Run 1 and 2 data.
For example, normalising by $B_c^+ \to J/\psi \overline{\mu} \nu_{\mu}$ would yield a constraint on the ratio $V_{cs}/V_{cb}$.
Due to CKM suppression, a measurement of $B_c^+ \to B^0 \overline{\mu} \nu_{\mu}$ is likely to require many more $B_c^+$ decays.

A lattice study of the $B_c^+ \to B_s^0 \overline{\ell} \nu_{\ell}$ and $B_c^+ \to B^0 \overline{\ell} \nu_{\ell}$ decays involves the practical complication of a heavy spectator quark. 
Care must be taken in placing such a particle on the lattice to avoid large discretisation effects. 
We consider two formalisms for the $b$ quark.
A valence NRQCD \cite{Lepage:1992tx,Dowdall:2011wh} $b$ quark, a formalism constructed from a non-relativistic effective theory, is used to simulate with physically massive $b$ quarks.
A complementary calculation uses HPQCD's heavy-HISQ method \cite{McNeile:2010ji,McNeile:2011ng,McLean:2019qcx}.
Here, all flavours of quark are implemented with the HISQ \cite{Follana:2006rc} formalism.
This is a fully relativistic approach which involves calculations for a set of quark masses on ensembles of lattices with a range of fine lattice spacings, enabling a fit from which the physical result at the $b$ quark mass in the continuum can be determined.
The method with an NRQCD bottom quark also uses HISQ for the charm, strange and down flavours.
This study will demonstrate the consistency of the NRQCD and heavy-HISQ approaches by comparing the form factors extrapolated to the physical-continuum limit.

In the limit of massless leptons, the differential decay rates for $B_c^+ \to B_s^0 \overline{\ell} \nu_{\ell}$ and $B_c^+ \to B^0 \overline{\ell} \nu_{\ell}$ are given by
\begin{align}
\frac{d \Gamma}{d q^2} = \frac{G_F^2 |V|^2}{24 \pi^3} |\mathbf{p}_2|^3 |f_+ (q^2)|^2,
\end{align}
where $V$ is the relevant associated CKM matrix element $V_{cs}$ or $V_{cd}$ and $f_+$ is one of two form factors that parametrise the continuum weak matrix element
\begin{align} \label{form factors}
\langle  B_{s(d)} (\mathbf{p}_2) & | V^{\mu} | B_c (\mathbf{p}_1) \rangle =  f_0 (q^2) \Bigg[ \frac{M_{B_c}^2 - M_{B_{s(d)}}^2}{q^2}q^{\mu} \Bigg] \nonumber \\
&+ f_+(q^2) \Bigg[ p_2^{\mu} + p_1^{\mu} - \frac{M_{B_c}^2 - M_{B_{s(d)}}^2}{q^2}q^{\mu} \Bigg] .
\end{align}
\begin{table}
\centering
\caption{Parameters for the MILC ensembles of gluon field configurations. The lattice spacing $a$ is determined from the Wilson flow parameter $w_0$~\cite{Borsanyi:2012zs} given in lattice units for each set in column 2 where values were obtained from \cite{Chakraborty:2016mwy} on sets 1 to 5 and \cite{Chakraborty:2014aca} on set 6. The physical value $w_0 = 0.1715(9)$ was fixed from $f_{\pi}$ in \cite{Dowdall:2013rya}. Sets 1 and 2 have $a \approx 0.15$ [fm], and sets 3 and 4 have $a \approx 0.12$ [fm]. Sets 5 and 6 have $a \approx 0.09$ [fm] and $a \approx 0.06$ [fm] respectively. Sets 1, 3, 5 and 6 have unphysically massive light quarks such that $m_l/m_s = 0.2$. Sets 1 to 5 were used in the NRQCD calculation of the form factors. The heavy-HISQ calculation used sets 3, 5 and 6.}
 \begin{tabular}{c c c c c c c} 
 \hline\hline
set & $w_0/a$ & $N_x^3 \times N_t$ & $n_\text{cfg}$ & $am_l^{\text{sea}}$ & $am_s^{\text{sea}}$ & $am_c^{\text{sea}}$ \\ [0.1ex] 
\hline
1 & 1.1119(10) & $16^3 \times 48$ & $1000$ & $0.013$ & $0.065$ & $0.838$\\
2 & 1.1367(5) & $32^3 \times 48$ & $500$ & $0.00235$ & $0.0647$ & $0.831$\\
3 & 1.3826(11) & $24^3 \times 64$ & $1053$ & $0.0102$ & $0.0509$ & $0.635$\\
4 & 1.4149(6) & $48^3 \times 64$ & $1000$ & $0.00184$ & $0.0507$ & $0.628$\\
5 & 1.9006(20) & $32^3 \times 96$ & $504$ & $0.0074$ & $0.037$ & $0.440$\\
6 & 2.896(6) & $48^3 \times 144$ & $250$ & $0.0048$ & $0.024$ & $0.286$\\
 \hline\hline
\end{tabular}
\label{LattDesc1}
\end{table}
\begin{table}
\caption{The HISQ valence quark masses for the light, strange and charm flavours for each of the sets described in Table \ref{LattDesc1}.
For the light quarks, the values for the valence quarks are identical to the masses of the light sea quarks.
The masses for the valence strange quarks and the valence charm quarks were tuned in \cite{Chakraborty:2014aca} using $w_0$ (Table \ref{LattDesc1}) to fix the lattice spacing. The fourth and fifth columns give the valence charm quark masses for the calculations with NRQCD and HISQ spectator quarks respectively. In the calculation with NRQCD spectator quarks slightly different $am_c^\text{val}$ values were used for historical reasons. Our fits allow for mistuning of the charm quark mass.}
\begin{tabular}{c c c c c}
\hline \hline
 & & & \multicolumn{2}{c}{$am_c^{\text{val}}$} \\
\multicolumn{1}{c}{} set & $am_l^{\text{val}}$ & $am_s^{\text{val}}$ & \multicolumn{1}{c}{NRQCD spectator} & \multicolumn{1}{c}{HISQ spectator} \\
\hline
1 & 0.013 & 0.0705 & 0.826 & \\
2 & 0.00235 & 0.0677 & 0.827 &  \\
3 & 0.0102 & 0.0541 & 0.645 & 0.663\\
4 & 0.00184 & 0.0507 & 0.631 &  \\
5 & 0.0074 & 0.0376 & 0.434 & 0.450\\
6 & 0.0048 & 0.0234 &  & 0.274\\            
\hline \hline                  
\end{tabular}
\label{tab:valencemass}
\end{table}
The 4-momentum transfer is $q= p_1 - p_2$, and only the vector part of the $V-A$ weak current contributes since QCD conserves parity.
The contribution of $f_0$ to the decay rate is suppressed by the lepton mass and hence irrelevant for the decays to $e \overline{\nu}_e$ and $\mu \overline{\nu}_{\mu}$.
The phase space is sufficiently small to disallow decays to $\tau \overline{\nu}_{\tau}$.
Form factors are constructed from the matrix elements that are obtained by fitting the appropriate lattice QCD 3-point correlator data.
By calculating correlators at a range of transfer momenta on lattices with different spacings and quark masses, continuum form factors at physical quark masses are obtained and then appropriately integrated to offer a direct comparison with decay rates that could be measured in experiment.

In this study, we begin with Sec.~\ref{sec:LattCalc} in which details of the lattice calculations are described.
Sec.~\ref{subsec:params} reports on the parameters and gauge configurations used to generate the propagators.
Next, Sec.~\ref{subsec:2ptcorrels} explains how the correlators are subsequently constructed for the two different treatments of the heavy spectator quark, as well as how the correlator data is fit to extract the matrix elements.
Our non-perturbative renormalisation method required to obtain the form factors is set out in Sec.~\ref{subsec:extract_ff}. Sec.~\ref{sec:results} presents results of the lattice calculations.
 Correlator fits are examined in Sec.~\ref{subsec:res_correlators}, whilst Sec.~\ref{subsec:res_ZV} discusses results for the renormalisation of the local lattice vector current.
 In Sec.~\ref{subsec:res_ff}, the form factor data for the cases of an NRQCD spectator and a HISQ spectator are plotted alongside.
Sec.~\ref{sec:discussion} is concerned with the methodology and results from fitting the form factor data.
An extrapolation of the form factors to physical-continuum point is presented in 
Sec.~\ref{sec:ff_chained_fit} and Sec.~\ref{sec:ff_spectator} shows 
how the form factors depend on the mass of the spectator quark. 
Finally, in Sec.~\ref{sec:conclusions} we give our conclusions.

\section{Lattice Calculation} \label{sec:LattCalc}

\subsection{Parameters and Set-up} \label{subsec:params}

We use ensembles with $2+1+1$ flavours of HISQ sea quark generated by the MILC Collaboration \cite{Bazavov:2010ru,Bazavov:2012xda,Bazavov:2015yea} and described in Table \ref{LattDesc1}.
The Symanzik-improved gluon action used is that from \cite{Hart:2008sq}, where the gluon action is improved perturbatively through $\mathcal{O}(\alpha_s)$ 
including the effect of dynamical HISQ sea quarks. The lattice spacing is identified by comparing the physical value for the Wilson flow parameter $\omega_0 = 0.1715(9)$ fm \cite{Dowdall:2013rya} with lattice values for $\omega_0 / a$ from \cite{Chakraborty:2016mwy} and \cite{Chakraborty:2014aca}.
Our calculations feature physically massive strange quarks and equal mass up and down quarks, with a mass denoted by $m_l$, with $m_l/m_s = 0.2$ and also the physical value $m_l/m_s = 1/27.4$ \cite{Bazavov:2014wgs}.
For sets 1 to 5 in Table \ref{LattDesc1}, strange propagators were re-used from \cite{Koponen:2017fvm}, a study of the pseudoscalar meson electromagnetic form factor.
Light propagators were re-used from \cite{Koponen:2015sgx}, an extension of \cite{Koponen:2017fvm} to the pion.
The valence quark masses used for the HISQ propagators on these gluon configurations are given in Table~\ref{tab:valencemass}.
The valence strange and charm quark masses used here were tuned 
in \cite{Chakraborty:2014aca, Koponen:2017fvm}, slightly away 
from the sea quark masses to yield results that more closely correspond to physical values.
The propagators were calculated using the MILC code~\cite{MILCgithub}. 

\begin{table}
\centering
\caption{The bottom quark masses, NRQCD action parameters $c_j$, and values for the tadpole improvement $u_0$ were obtained from \cite{Dowdall:2011wh}.
The final columns gives the different momenta for the strange and light quarks considered in the NRQCD calculation implemented with twisted boundary conditions.}
\begin{tabular}{ c c c c c c c c c c c c} 
\hline\hline
set &  $am_b^{\text{val}}$ & $c_1,c_6$ & $c_5$ & $c_4$ & $u_0$ & $|a\mathbf{q}|$ \\ [0.1ex]
\hline
1 &  $3.297$ & $1.36$ & $1.21$ & $1.22$ & $0.8195$ & 0& 0.1243& 0.3730& 0.6217 \\
2 &  $3.25$ & $1.36$ & $1.21$ & $1.22$ & $0.8195$ & 0& 0.3649 \\
3 &  2.66 & 1.31 & 1.16 & 1.20 & 0.8341& 0&0.1& 0.3& 0.5\\
4 & $2.62$ & $1.31$ & $1.16$ & $1.20$ & $0.8341$ & 0 \\
5 &  $1.91$ & $1.21$ & $1.12$ & $1.16$ & $0.8525$ & 0& 0.0728& 0.364& 0.437 \\
 \hline\hline
\end{tabular}
\label{LattDesc2}
\end{table}

\begin{table}
\centering
\caption{Heavy quark masses and momenta used for the heavy-HISQ calculation.
The momenta are in the $(1\hspace{1mm}1\hspace{1mm}1)$ direction.}
\begin{tabular}{c c c c c c c c c c c c c c} 
\hline\hline
set & $am_h^{\text{val}}$ &  &  & & $|a\mathbf{q}|$ \\ [0.1ex] 
\hline
3 &  0.663 & 0.8 & & &0&0.1&0.3&0.5 \\
5  &0.450&0.6&0.8& &0&0.07281&0.218&0.364&0.437 \\
6 &0.274&0.450&0.6&0.8&0&0.143&0.239&0.334 \\
 \hline\hline
\end{tabular}
\label{LattDescHHISQmassesmom}
\end{table}

We work in the frame where the $B_c^+$ is at rest, and momentum is inserted into the strange or down valence quark through 
twisted boundary conditions \cite{Sachrajda:2004mi, Guadagnoli:2005be} 
in the $(1\hspace{1mm}1\hspace{1mm}1)$ direction.
The values of the momenta used are given in Tables~\ref{LattDesc2} 
and~\ref{LattDescHHISQmassesmom}.
The periodic boundary conditions of the fermion fields are modified by phases $\theta_i$
\begin{align}
\psi (n + N_x \hat{i}) &= e^{i \pi \theta_i} \psi(n)
\end{align}
so that the usual lattice momenta $q_i = 2\pi k_i/aN_x$, for integers $k_i$, are shifted by $\pi \theta_i/aN_x$.
The corresponding $q^2$ is then constructed by taking $q_0$ to be the difference in energies of the lowest lying initial and final states.

The coefficients of operators corresponding to 
relativistic correction terms in the NRQCD action 
are given in Table \ref{LattDesc2}.
The valence $b$ quark masses used for the NRQCD propagators are 
also given there. 
The values were taken from \cite{Dowdall:2011wh}, where the $b$ quark mass was found by matching the experimental value for the spin-averaged kinetic mass of the $\Upsilon$ and the $\eta_b$ to lattice data.
For the calculation with an NRQCD spectator bottom quark, we use sets 1 to 5 in Table \ref{LattDesc1}.

Bare heavy quark masses $am_h$ used for the heavy-HISQ method are shown in Table \ref{LattDescHHISQmassesmom}.
The selection of heavy quark masses follows \cite{McLean:2019qcx}.
As well as sets 3 and 5, the heavy-HISQ calculation makes use of a lattice finer than the five sets featuring in the calculation with an NRQCD spectator, set 6 in Table \ref{LattDesc1}.
This is motivated by the necessity to avoid large discretisation effects 
that grow with $(am_h)$ (as $(am_h)^4$ at tree-level) whilst gathering data at large masses that will reliably inform the limit $m_h \to m_b$.

\subsection{Correlators} \label{subsec:2ptcorrels}
\subsubsection{NRQCD spectator case}
For the case of an NRQCD spectator quark, random wall source \cite{Aubin:2004fs} HISQ propagators with the mass of the charm quark are calculated and combined with random wall source NRQCD $b$ propagators to generate $B_c^+$ 2-point correlator data.
2-point correlators for $B_{s(d)}^0$ are generated similarly.
The strategy of combining NRQCD random wall propagators and HISQ random wall propagators to yield 2-point correlators was first developed in \cite{Gregory:2010gm}.
NRQCD propagators are generated by solving an initial value problem.
This is computationally very fast compared to calculating rows of the 
inverse of the quark matrix. 

\begin{figure}[t]
\centering
%
\includegraphics[width=0.5\textwidth]{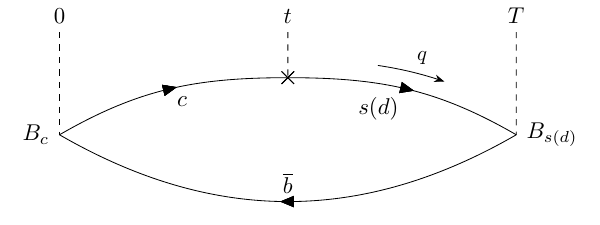}
\caption{3-point correlator $C_{\text{3pt}} (t,T)$. The flavour-changing operator insertion is denoted by a cross at timeslice $t$ and the total time length of the 3-point correlator is $T$. The random wall source for the $b$ and $s/d$ propagators is at the timeslice of the $B_{s(d)}$ interpolator.}
\label{3ptcorrelfig}
\end{figure}

The 3-point correlator needed here is represented diagrammatically in Fig.~\ref{3ptcorrelfig}. A HISQ charm quark propagator is generated by using the random wall bottom quark propagator as a sequential source.
Following Appendix B in~\cite{Donald:2012ga}, 
and excluding a spacetime-dependent 
sign, the sequential source is given by the spin-trace 
\begin{align} \label{NRQCDcharmseqsrc}
\text{Tr}_{\text{spin}} \Big \{ \Gamma \Omega^{\dagger}(\mathbf{x},0) S_b^{\text{RW}} (\mathbf{x},0) \Big \},
\end{align}
where $\Gamma$ is the gamma matrix structure at the operator insertion, $S_b^{\text{RW}}$ is the random wall NRQCD propagator, and 
\begin{align}
\Omega(x) \coloneqq& \prod_{\mu=1}^4 (\gamma_{\mu})^{\frac{x_{\mu}}{a}}
\end{align}
is the space-spin matrix which transforms the naive quark field to diagonalise the HISQ action in spin-space.

\subsubsection{HISQ spectator case}
\begin{figure}[t] 
\centering
\includegraphics[width=0.45\textwidth]{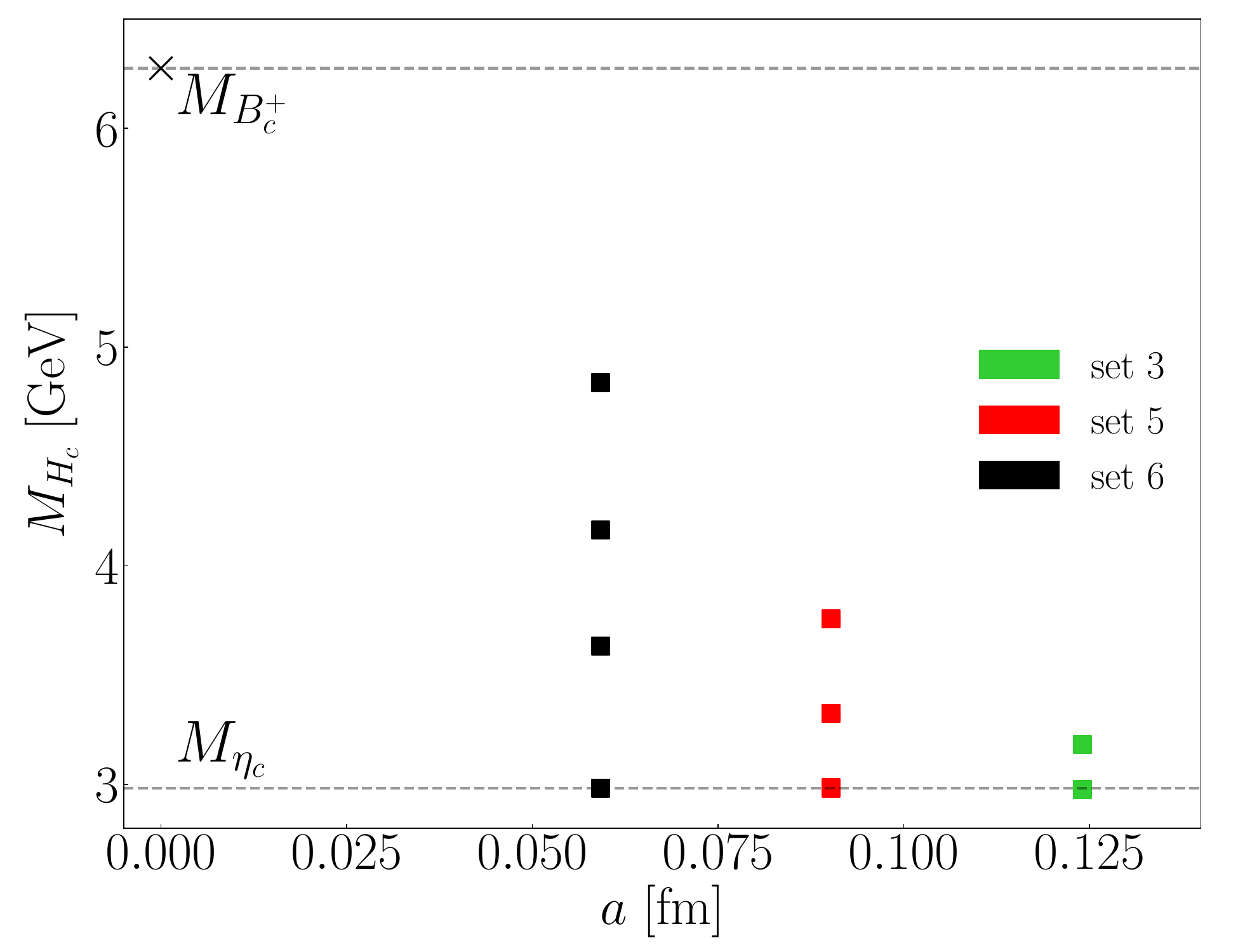}
\caption{The mass $M_{H_c}$ of the heavy-charm meson is plotted against lattice spacing for each of the values of $am_h$ used in the heavy-HISQ calculation. Obtained from fitting the correlators as described in Eq.~(\ref{corrfitform}), $M_{H_c}$ is a proxy for the bare lattice heavy quark mass $am_h$. The continuum-physical point is denoted by a cross at $a=0$ [fm] and $M_{H_c} = M_{B_c^+}$.
Note that the $y$-axis scale begins near $M_{\eta_c}$.}
\label{hhisqmassplot}
\end{figure}

The case of a HISQ spectator quark proceeds similarly with the only difference being the use of a HISQ propagator instead of an NRQCD propagator for the bottom quark.
Again, the charm propagator uses the spectator bottom quark propagator as a sequential source.
Multiple masses are used for the spectator quark, each requiring a different charm propagator for the 3-point correlator.
Fig.~\ref{hhisqmassplot} shows the heavy-charm pseudoscalar meson masses that arise from calculations with the $am_h$ values in Table \ref{LattDescHHISQmassesmom}.
On set 6, the finest lattice considered, we reach a value for $M_{H_c}$ that is $80\%$ of the physical $B_c$ mass.

The same strange and light random wall HISQ propagators on sets 3 and 5 are used in both the NRQCD and the heavy-HISQ calculations, thus the data on these lattices in the two approaches will be correlated. However, the effect of these correlations is small in the physical-continuum limit since the heavy-HISQ data on sets 3 and 5 are the furthest away from the physical $b$ quark mass point, and hence these correlations are safely ignored.

\subsubsection{Fitting the correlators} \label{2ptcorrels}
The correlator fits minimise an augmented $\chi^2$ function 
as described in \cite{Lepage:2001ym,Hornbostel:2011hu,Bouchard:2014ypa}.
The functional forms for the 2-point and 3-point correlators 
\begin{align} \label{corrfitform}
C^{B_{s(d)}}_{\text{2pt}} (t) & = \sum_i a[i]^2 e^{-E_a[i] t} - \sum_i a_o[i]^2 (-1)^t e^{-E_{a_o}[i]t} \nonumber \\
C^{B_c}_{\text{2pt}} (t) & = \sum_j b[j]^2 e^{-E_b[j] t} - \sum_j b_o[j]^2 (-1)^t e^{-E_{b_o}[j]t} \nonumber \\
C_{\text{3pt}} (t,T) & = \sum_{i,j} a[i] e^{-E_a[i]t}  V_{nn}[i,j] b[j] e^{-E_b[j](T-t)} \nonumber \\
- \sum_{i,j} & (-1)^{T-t} a[i] e^{-E_a[i]t}  V_{no}[i,j] b_o[j] e^{-E_{b_o}[j](T-t)} \nonumber \\
- \sum_{i,j} & (-1)^t a_o[i] e^{-E_{a_o}[i]t}  V_{on}[i,j] b[j] e^{-E_b[j](T-t)} \nonumber \\
+ \sum_{i,j} & (-1)^T a_{o}[i] e^{-E_{a_o}[i]t}  V_{oo}[i,j] b_o[j] e^{-E_{b_o}[j](T-t)} \nonumber \\
\end{align}
follow from their spectral decomposition and include oscillatory contributions from the staggered quark time-doubler.
The matrix elements are related to the fit parameters $V_{nn}[i,j]$ through
\begin{align} \label{Jopinsert}
V_{nn} [0,0] = \frac{\langle B_{s(d)} | J | B_c \rangle}{\sqrt{2E_{B_{s(d)}} 2E_{B_c}}},
\end{align}
where $J$ is the relevant operator that facilitates the $c \to s(d)$ flavour transition.
The pseudoscalar mesons of interest are the lowest lying states consistent with their quark content, so we are only concerned with the matrix elements for $i=j=0$ since we restrict $E[k] \leq E[k+1]$ by using log-normal prior distributions for the energy differences.
The presence of $i,j>0$ terms are necessary to give a good fit and to allow for the full systematic uncertainty from the presence of excited states to be included in the extracted $V_{nn} [0,0]$.
On each set, the 2-point and 3-point correlator data for both $c \to s$ and $c \to d$ at all momenta are fit simultaneously to account for all possible correlations. The matrix elements and energies are extracted and form factor 
values determined, along with the correlations between results at different 
momenta.


\subsection{Extracting The Form Factors} \label{subsec:extract_ff}
The Partially Conserved Vector Current (PCVC) Ward identity allows for a fully non-perturbative renormalisation of the lattice vector current.
Since the same HISQ action is used for the $c$ and $s(d)$ quarks that couple to the $W^+$ in both the NRQCD and heavy-HISQ approaches, we have the PCVC identity
\begin{align}\ \label{contPCVC}
\partial_{\mu} V^{\mu} _{\text{cons}}= (m_c - m_{s(d)}) S,
\end{align}
relating the conserved (point-split) $c \to s(d)$ lattice vector current and the local lattice scalar density $S$.
We choose a local lattice operator $V^{\mu}_{\text{loc}}$, thus Eq.~(\ref{contPCVC}) must be adjusted by a single renormalisation factor $Z_V$ associated with that operator, giving
\begin{align}
q_{\mu} \langle B_{s(d)} | V^{\mu}_{\text{loc}} | B_c \rangle Z_V = (m_c - m_{s(d)})  \langle B_{s(d)} | S | B_c \rangle.\label{PCVClatt}
\end{align}
Since $Z_V$ is $q^2$ independent, in principle $Z_V$ need only 
be found at zero-recoil where $q^{\mu}$ has only a temporal component~\cite{Koponen:2013tua}. 
This avoids the need to calculate 3-point correlators associated with the spatial components of the vector current matrix element that appear in Eq.~(\ref{PCVClatt}) for $\mathbf{q} \neq \mathbf{0}$.
However, in practice, it is preferable to determine $f_+$ near zero-recoil through the spatial components of the vector current matrix element, albeit with the additional cost in computing 3-point correlators with the corresponding insertion.

As in~\cite{Na:2010uf}, we combine Eqs.~(\ref{form factors}) and (\ref{PCVClatt}) to give a determination 
\begin{align}
f_0 \big(q^2 \big) = \langle B_{s(d)} | S | B_c \rangle \frac{m_c - m_{s(d)}}{M_{B_c}^2 - M_{B_{s(d)}}^2} \label{f0scalar}
\end{align}
of $f_0$  solely in terms of the scalar density matrix element. 
We use Eq.~(\ref{f0scalar}) and calculation of the vector current matrix element to determine $f_+$ and $f_0$ for the full $q^2$ range following 
~\cite{Koponen:2013tua, Chakraborty:2017pud}.
Thus, we will calculate matrix elements of both the local scalar density $J=S$ and the local vector current $J=V$.

Once $f_0$ is determined, $f_+$ is obtained using Eq.~(\ref{form factors}) for $\mu = 0$ to yield
 \begin{align} \label{fplusextract}
f_+(q^2) = \frac{ Z_V \mathcal{V}^{0} - q^0 f_0(q^2)\frac{M_{B_c^+}^2 - M_{B_s^0}^2}{q^2}}{ p_2^0 + p_1^0 - q^0 \frac{M_{B_c^+}^2 - M_{B_s^0}^2}{q^2}},
 \end{align}
where $\mathcal{V}^\mu$ is the vector current matrix element, except at zero-recoil where the denominator vanishes and $f_+$ cannot be extracted.
We find that using Eq.~(\ref{fplusextract}) near zero-recoil is problematic since both the numerator and denominator grow from 0 as $q^2$ is decreased from the maximum value at zero-recoil.
For the case where the spectator is an NRQCD $b$ quark, we instead use Eq.~(\ref{form factors}) with $\mu = i \neq 0$
\begin{align}
f_+ (q^2) &= \frac{- \frac{Z_V \mathcal{V}^{i}}{q^i} + f_0(q^2)\frac{M_{B_c^+}^2 - M_{B_s^0}^2}{q^2}}{ 1 + \frac{M_{B_c^+}^2 - M_{B_s^0}^2}{q^2}} .\label{fplusextract2}
\end{align}
This method gives much smaller errors near to zero-recoil.
Although mathematically equivalent to Eq.~(\ref{fplusextract}), extracting $f_+$ through Eq.~(\ref{fplusextract2}) does not suffer an inflation of error near zero-recoil since both the numerator and denominator are non-zero for all physical $q^2$.
However, since $\mathcal{V}^i$ appears explicitly in Eq.~(\ref{fplusextract2}), 3-point correlators with an insertion of $V^i$ need to be calculated.
For the case of the spectator NRQCD $b$ quark, the use of Eq.~(\ref{fplusextract2}) is straightforward except that it requires inversions of the charm quark propagator from a different sequential source (see Eq.~(\ref{NRQCDcharmseqsrc})) to allow for insertion of the current $V^i = \gamma^i \otimes \gamma^i$ in the mixed NRQCD-HISQ 3-point function.
Collecting $\mathcal{V}^i$ at non-zero 3-momentum transfer in the NRQCD calculation will also test for any $q^2$ dependence of $Z_V$ that would appear as a discretisation effect.

Using Eqs.~(\ref{f0scalar}) and (\ref{fplusextract}) or (\ref{fplusextract2}), form factor data at a variety of lattice spacings, light quark masses and momenta are obtained from the energies and matrix elements. 

\subsubsection{NRQCD spectator case}
For the case of an NRQCD spectator quark, the form factor extraction is complicated by the energy offset as a consequence of the subtraction of the $b$ quark rest mass inherent in the NRQCD formalism.
Whilst physical energy differences are preserved with NRQCD quarks, energy sums are not.
Consequently, Particle Data Group (PDG) \cite{PDG} values are used where necessary. For example, we take
\begin{align}
M_{B_c}^2 - M_{B_s(d)}^2 &= \Big( E^{\text{sim}}_{B_c} (|a\mathbf{q}| = 0) - E^{\text{sim}}_{B_{s(d)}} (|a\mathbf{q}| = 0)  \Big) \nonumber \\
& \hspace{15mm} \times \Big( M_{B_c}^{\text{PDG}} + M_{B_s(d)}^{\text{PDG}} \Big)
\end{align}
when extracting the form factors.
We use interpolating operators $\overline{c} \gamma_5 b$ and $\overline{s} \gamma_5 b$ ($\overline{d} \gamma_5 b$) for $J^P = 0^-$ pseudoscalars $B_c^+$ and $B_{s(d)}^0$ respectively.
\\

\subsubsection{HISQ spectator case}
For the case of a HISQ spectator quark, we work only with local scalar and vector currents.
Expressed in the spin-taste basis, we use $\gamma_5 \otimes \gamma_5$ for the $H_{s(d)}$ interpolating operator and two different operators, 
$\gamma_5 \otimes \gamma_5$ and $\gamma_5 \gamma_0 \otimes \gamma_5 \gamma_0$, for the $H_c$ interpolator.
The first of these, $\gamma_5 \otimes \gamma_5$, makes a tasteless 3-point correlation function when the scalar density operator $\mathbb{1} \otimes \mathbb{1}$ is used.
The second, $\gamma_5 \gamma_0 \otimes \gamma_5 \gamma_0$, allows for a tasteless 3-point correlation function when we use the local temporal vector current operator $\gamma_0 \otimes \gamma_0$ \cite{Koponen:2013tua}.
This requires the calculation of two $H_c$ 2-point functions with the two different choices of operator at both the source and the sink.
The difference in masses between these two different tastes of $H_c$ meson is tiny and, although consistently taken care of, it has no impact on the calculation.

\section{Results} \label{sec:results}
\subsection{Correlators} \label{subsec:res_correlators}

\begin{figure}[t] 
\centering
\includegraphics[width=0.45\textwidth]{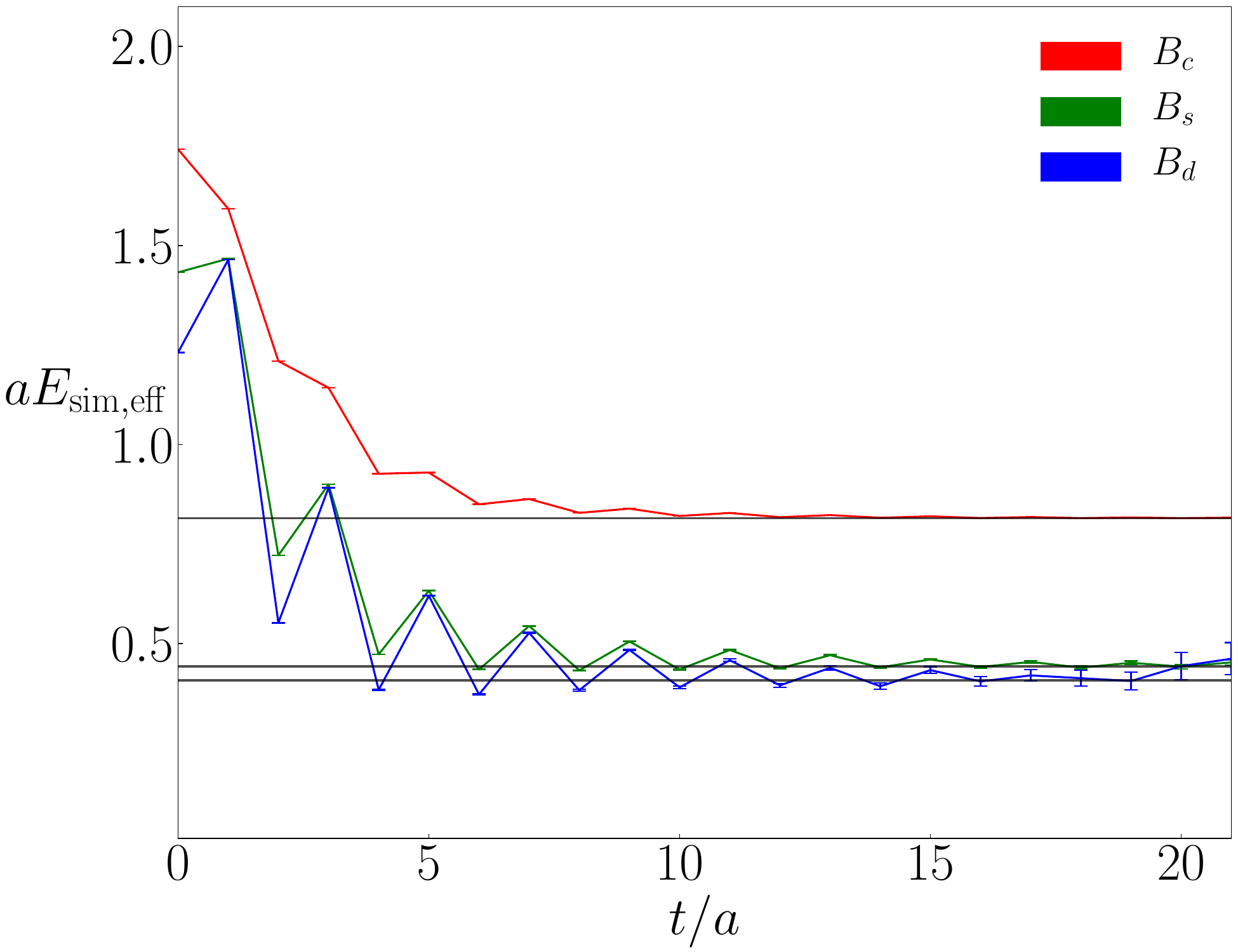}
\caption{Effective simulation energies (Eq.~(\ref{effsimenergy})) of 2-point correlators with an NRQCD spectator quark on set 5 for $|a\mathbf{q}| = 0.427$. The black lines with grey error bands show the energies extracted from fitting the correlators in the simultaneous fit of fine lattice data with all the 3-point correlators and all the momenta. The $B$ meson energies shown here are offset from their physical values as a consequence applying the NRQCD formalism to the constituent $b$ quark.}
\label{simeffenergies}
\end{figure}

\begin{figure}[t] 
\centering
\includegraphics[width=0.45\textwidth]{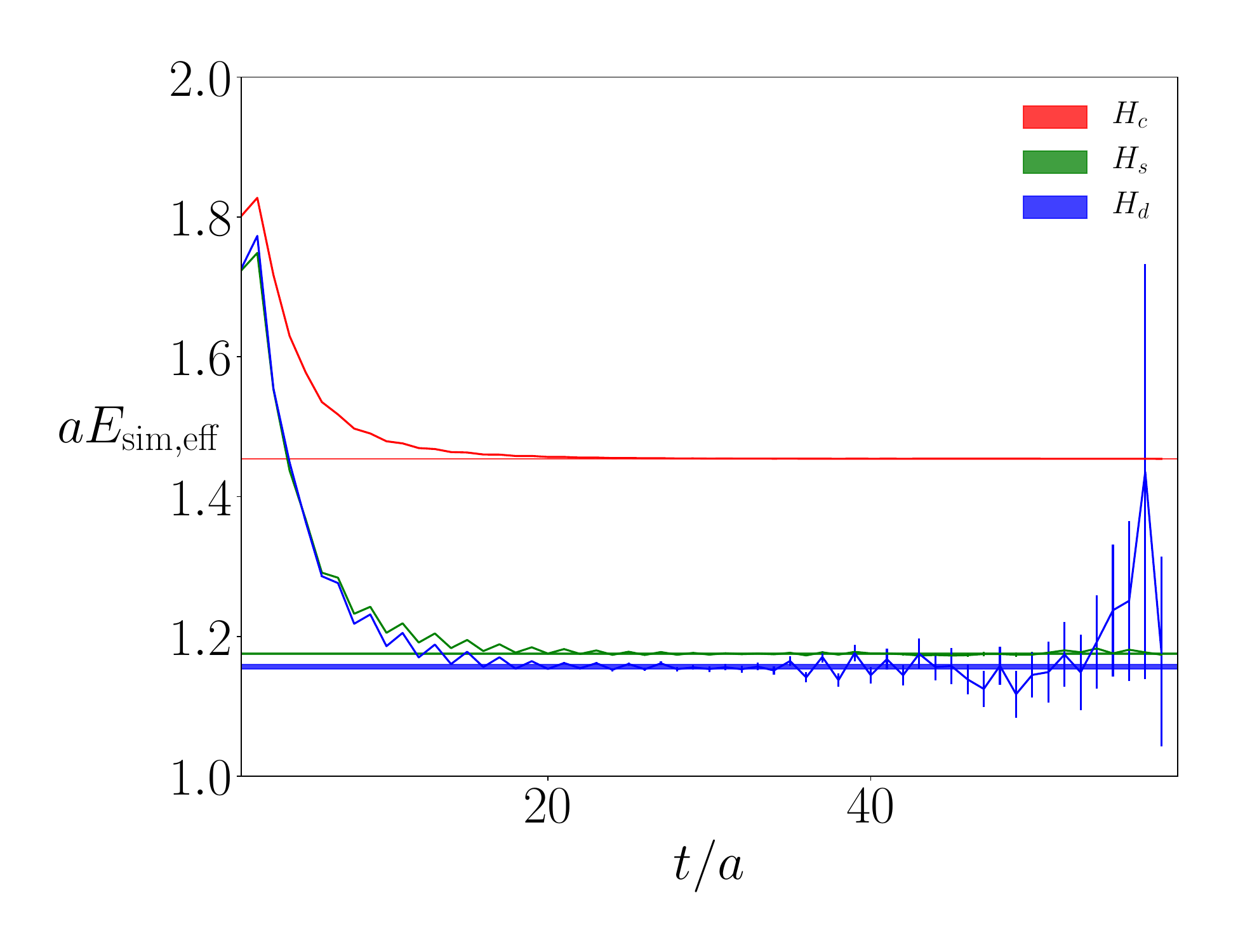}
\caption{Effective simulation energies (Eq.~(\ref{effsimenergy})) of 2-point correlators with a HISQ spectator quark on set 6 at zero twist with $am_h=0.8$. The horizontal bands show the energies extracted from the full simultaneous correlator fit.}
\label{simeffenergiesHeavyHISQ}
\end{figure}

Figs.~\ref{simeffenergies} and \ref{simeffenergiesHeavyHISQ} provide samples of the correlator data from the NRQCD and heavy-HISQ calculations respectively. 
The quantity plotted is the effective simulation energies, which we define by the two-step log-ratio
\begin{align}
aE_{\text{sim,eff}} &= \frac{1}{2} \log \Big( \frac{C(t)}{C(t+2)} \Big). \label{effsimenergy}
\end{align}
This ratio is preferable to an effective energy defined using $C(t)/C(t+1)$ since the ratio in Eq.~(\ref{effsimenergy}) better suppresses the oscillatory contributions in Eq.~(\ref{corrfitform}).
Error bars are present in the figure but mostly too small to observe.
We exclude $t_\text{min}/a$ data points from the beginning and end points of the correlators in our fits to reduce the contributions from excited states.

For each of the cases of an NRQCD and HISQ spectator quark, we fit all of the correlator data to Eq.~(\ref{corrfitform}) on each set simultaneously to obtain the correlations between the fitted parameters.
Consequently, the correlator fits involve a large covariance matrix.
Without extremely large statistical samples of results 
small eigenvalues of the covariance matrix are underestimated~\cite{Michael:1993yj, Dowdall:2019bea} and this 
causes problems when carrying out the inversion to find $\chi^2$.
We overcome this by using an SVD (singular-value decomposition) cut; any eigenvalue of the covariance matrix smaller than some proportion $c$ of the biggest eigenvalue $\lambda_{\text{max}}$ is replaced by $c\lambda_{\text{max}}$.
By carrying out this procedure, the covariance matrix becomes less singular.
These eigenvalue replacements will only inflate our final errors, hence this strategy is conservative.
The SVD cut reduces the $\chi^2 /$d.o.f. reported by the fit 
because it lowers the contribution to $\chi^2$ of the modes with eigenvalues below the 
SVD cut. 
In order to check the suitability of the SVD cut,
we must test the goodness-of-fit 
from a fit where noise (SVD-noise) is added to the data to reinstate the size of 
fluctuations expected from the modes below SVD cut, as described in Appendix D of~\cite{Dowdall:2019bea}.
The $\chi^2_{\text{SVD-noise}} / $d.o.f. is used to check the goodness of fit for both cases of spectator quark.

Many fits were carried out with different SVD cuts, number of exponentials $N$, and positions $t_{\text{min}}^{\text{2pt}}$ and $t_{\text{min}}^{\text{3pt}}$ of the first timeslice where the correlators are fit.
We selected the fit of the correlators on each lattice for form factor extraction based on the $\chi^2 / \text{d.o.f.}$ and $Q$-value.

The parameters used in the fits of correlators with an NRQCD spectator quark are presented in Table \ref{tab:NRQCDparams}. The parameters given in bold are those used for 
our final fits. Other values are used in tests of the stability of our form factor fits 
to be discussed in Sec.~\ref{sec:NRQCDFFF}. 

\begin{table}
\centering
\caption{Input parameters (see text for definition) to the correlator fits for the calculation with NRQCD spectator quarks together with fits including variations of $t_\text{min}/a$, $N$ and SVD cut. Bold entries indicate those fits used to obtain our final result. Other values are used in tests 
of the stability of our form factor fits to be discussed in Sec.~\ref{sec:NRQCDFFF}.}
\begin{tabular}{c c c c c c} 
 \hline\hline
set & SVD cut & $t_\text{min}^\text{2pt}/a$  & $t_\text{min}^\text{3pt}/a$ & $N$ & $\chi^2_\text{SVD-noise}/\text{dof}$\\ [0.1ex] 
\hline
1 & \textbf{0.1} & \textbf{2} & \textbf{2} & \textbf{6} & \textbf{1.00}\\
 & 0.1 & 2 & 3 & 4 & 1.00\\
\hline
2 & \textbf{0.075} & \textbf{6} & \textbf{2} & \textbf{6} & \textbf{1.00}\\
 & 0.075 & 6 & 2 & 5 & 1.10\\
\hline
3 & \textbf{0.1} & \textbf{6} & \textbf{3} & \textbf{6} &  \textbf{1.00}\\
 & 0.075 & 4 & 2 & 6 & 1.00\\
\hline
4 & \textbf{0.025} & \textbf{4} & \textbf{3} & \textbf{6} & \textbf{1.00}\\
 & 0.075 & 4 & 2 & 6 & 0.95\\
\hline
5 & \textbf{0.05} & \textbf{6} & \textbf{2} & \textbf{6} & \textbf{1.00}\\
 & 0.3 & 4 & 3 & 6 & 1.00\\
 \hline \hline
\end{tabular}
\label{tab:NRQCDparams}
\end{table}

\begin{table}
\centering
\caption{Input parameters (see text for definition) to the heavy-HISQ correlator fits together with fits including variations of $t_\text{min}/a$, $N$ and SVD cut. Bold entries indicate those fits used to obtain our final result. Other values will be used in tests of the stability of our form factor 
fits in Sec.~\ref{subsec:hHISQ_ff_fits}. }
\begin{tabular}{c c c c c c} 
 \hline\hline
set & SVD cut & $t_\text{min}^\text{2pt}/a$  & $t_\text{min}^\text{3pt}/a$ & $N$& $\chi^2_\text{SVD-noise}/\text{dof}$\\ [0.1ex] 
\hline
3 & \textbf{0.025} & \textbf{6} & \textbf{2} & \textbf{4} & \textbf{0.94}\\
 & 0.025 & 6 & 2 & 3 & 1.05\\
 & 0.075 & 6 & 2 & 4 & 0.90\\
\hline
5 & \textbf{0.025} & \textbf{4} & \textbf{2} & \textbf{4} & \textbf{0.95}\\
 & 0.025 & 4 & 2 & 3 & 0.94\\
 & 0.075 & 4 & 2 & 4 & 0.96\\
\hline
6 & \textbf{0.025} & \textbf{6} & \textbf{3} & \textbf{4} & \textbf{0.95}\\
 & 0.025 & 4 & 2 & 3 & 0.99\\
 & 0.05 & 6 & 3 & 4 & 0.95\\
 \hline \hline
\end{tabular}
\label{HHparams}
\end{table}

We fit the heavy-HISQ correlator data to Eq.~(\ref{corrfitform}) on each set simultaneously, including correlations between data with different values of twist, heavy quark mass, and $H_{s/d}$ final state.
Values for $t_\text{min}/a$, the chosen SVD cut, the number of exponentials used in Eq.~(\ref{corrfitform}) and the resultant value of $\chi^2/\text{dof}$ including SVD noise are given in Table \ref{HHparams}.
We also include in Table \ref{HHparams} fits using variations of these parameters. Form factor fit coefficients obtained using combinations of these variations are shown in Figs.~\ref{stabplothh1} and \ref{stabplothh2} in Sec.~\ref{subsec:hHISQ_ff_fits} and demonstrate that our results are insensitive to such choices.

\subsection{Vector Current Renormalisation $Z_V$} \label{subsec:res_ZV}
In this section we give our results for the renormalisation factor $Z_V$ for the vector current (Eq.~(\ref{PCVClatt})) and test for dependence of $Z_V$ on $q^2$ (for the case of an NRQCD spectator) and on the spectator quark mass (for the case of a HISQ spectator).

\begin{figure}[t] 
\centering
\includegraphics[width=0.45\textwidth]{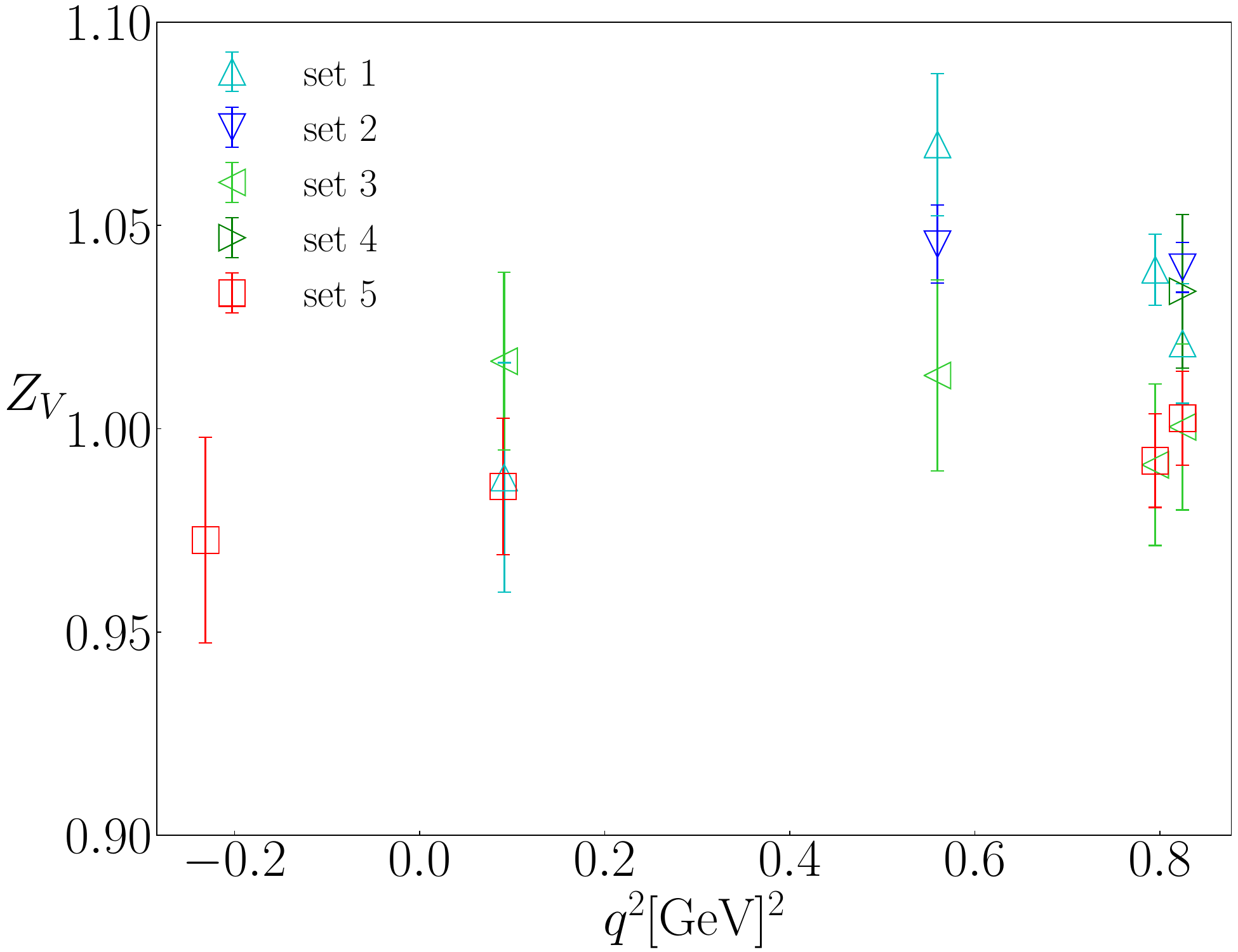}
\caption{$Z_V$ for the $c \to s$ vector current evaluated at different $q^2$ from the calculation with an NRQCD spectator quark using Eq.~(\ref{PCVClatt}).}
\label{Zq2}
\end{figure}

The vector current renormalisation factor $Z_V$ computed at different momentum transfer with NRQCD $b$ quarks shows no significant dependence on $q^2$ on each set, demonstrated by Fig.~\ref{Zq2}.
Mild lattice spacing dependence is observed, however.
For each momenta, we use the $Z_V$ found at the corresponding $q^2$ from Eq.~(\ref{PCVClatt}).

\begin{table}
\centering
\caption{$Z_V$ obtained at zero-recoil using an NRQCD spectator $b$ quark .}
\begin{tabular}{c c c} 
 \hline\hline
set &  $c \to s$ & $c \to d$ \\ [0.1ex] 
\hline
1 & 1.021(15) & 1.041(18) \\
2 & 1.0397(61) & 1.021(17) \\
3 & 1.000(20) & 1.004(22) \\
4 & 1.034(19) & 0.983(20) \\
5 & 1.003(12) & 0.958(20) \\
 \hline\hline
\end{tabular}
\label{Zcstab}
\end{table}

\begin{table}
\centering
\caption{$Z_V$ for $c \to s$ obtained at zero-recoil using a HISQ spectator quark with different values of the heavy quark mass $m_h$.}
\begin{tabular}{c c c c c c} 
 \hline\hline
set/$am_h$ & $0.274$ & $0.450$ & $0.6$ & $0.663$ & $0.8$ \\ [0.1ex] 
\hline
3 &  &  &  & 1.026(32) & 1.029(36) \\
5 &  & 1.006(17) & 1.003(19) &  & 1.000(20) \\
6 & 0.997(14) & 0.994(17) & 0.995(19) &  & 0.995(22) \\
 \hline\hline
\end{tabular}
\label{Zcstabheavy-HISQ}
\end{table}

\begin{table}
\centering
\caption{$Z_V$ for $c \to d$ obtained at zero-recoil using a HISQ spectator quark with different values of the heavy quark mass $m_h$.}
\begin{tabular}{c c c c c c} 
 \hline\hline
set/$am_h$ & $0.274$ & $0.450$ & $0.6$ & $0.663$ & $0.8$ \\ [0.1ex] 
\hline
3 &  &  &  & 1.016(47) & 1.019(50) \\
5 &  & 1.009(23) & 1.004(25) &  & 1.000(27) \\
6 & 0.996(22) & 0.993(25) & 0.994(28) &  & 0.995(32) \\
 \hline\hline
\end{tabular}
\label{Zcdtabheavy-HISQ}
\end{table}

\begin{figure}[t] 
\centering
\includegraphics[width=0.45\textwidth]{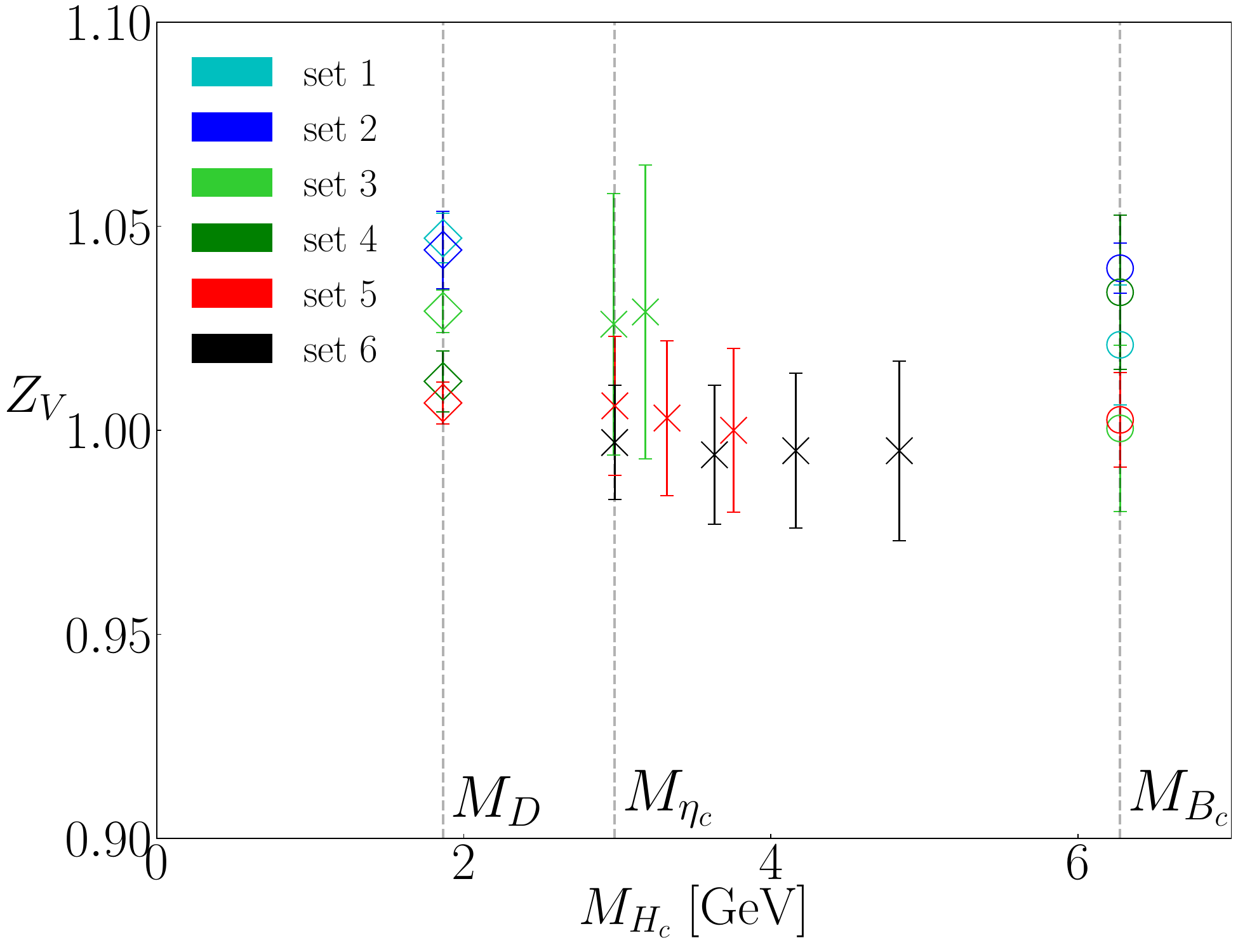}
\caption[]{$Z_V$ of the $c \to s$ vector current from both the NRQCD and heavy-HISQ calculations are plotted alongside values from $D \to K$ \cite{Koponen:2013tua}. The NRQCD data is marked with circles, the heavy-HISQ data is marked with crosses, and finally the $D \to K$ values are given by diamonds. As expected, no significant dependence on the spectator mass is observed.}
\label{Zctos_spec_mass}
\end{figure}

\begin{figure}[t] 
\centering
\includegraphics[width=0.45\textwidth]{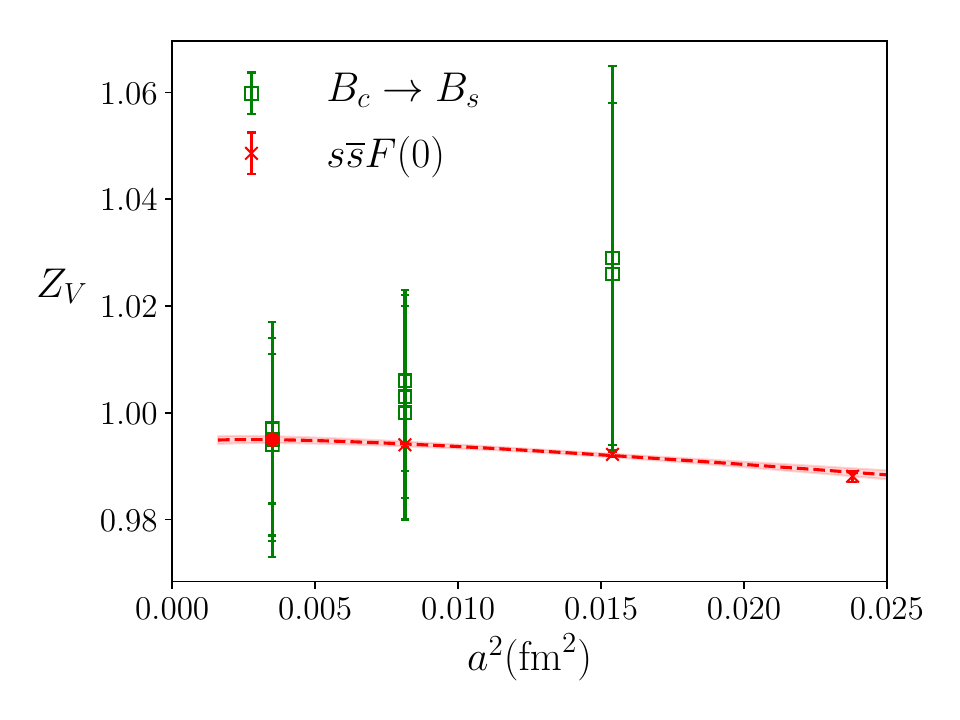}
\caption{$Z_V$ from the $s \to s$ vector current from~\cite{Chakraborty:2017hry} (red crosses) and the $c \to s$ vector current from the 
heavy-HISQ calculation given here (green squares). 
The curve is the fitted perturbative expansion, including discretisation effects, detailed in \cite{Chakraborty:2017hry}. The red circle is an extrapolated value at the lattice spacing associated with the superfine lattice.
}
\label{Zss}
\end{figure}

The $Z_V$ factor in Eq.~(\ref{PCVClatt}) is associated only with the local vector current operator and should be independent of the spectator quark. 
$Z_V$ values obtained in the different calculations are tabulated in 
Tables~\ref{Zcstab}, \ref{Zcstabheavy-HISQ} and~\ref{Zcdtabheavy-HISQ}. 
Good agreement is seen on set 5 at zero-recoil between the 
results with NRQCD and heavy-HISQ spectator quarks. 
Dependence on the mass of the spectator quark is displayed in 
Fig.~\ref{Zctos_spec_mass}. The plot includes values from the 
analogous calculation for the $D \to K$ case~\cite{Koponen:2013tua}.
For $D \to K$, a charm quark decays into a strange quark, as in $B_c \to B_s$, but here the spectator quark is a light quark, much less massive than the heavy spectator quark in $B_c \to B_s$.
The $Z_V$ from $B_c \to B_s$ and $D \to K$ in Fig.~\ref{Zctos_spec_mass} are 
nevertheless in good agreement, demonstrating negligible dependence on the mass of the quark spectating the $c \to s$ transition.

It is also of interest to compare vector current renormalisation 
factors for different masses of quark featuring in the current.
For example, \cite{Chakraborty:2017hry} calculates the local $\overline{s} \gamma_{\mu} s$ vector current renormalisation factor from an 
$\eta_s \to \eta_s$ 3-point correlation function at $q^2=0$ 
on the 2+1+1 MILC ensembles.
This gave very precise values and it was possible to fit $Z_V$ to a perturbative expansion in $\alpha_s$ (including the known first-order term) along with 
discretisation effects.
The fit is plotted in Fig.~\ref{Zss} alongside $Z_V$ for $c \to s$ values determined in this study.
This plot shows differing behaviour as a function of $a^2$.
The value for $Z_V$, determined non-perturbatively, is a combination of the underlying perturbative series in $\alpha_s$ evaluated at a scale related to the lattice spacing and discretisation effects that depend on how it was determined.
Since the underlying perturbative series is common to different determinations, comparison will reveal the differing discretisation effects.
Fig.~\ref{Zss} shows this in the comparison of our $Z_V$ values for the local $\overline{s} \gamma_{\mu} c$ current with those determined for the local $\overline{s} \gamma_{\mu} s$ current.
In the limit of vanishing lattice spacing, where discretisation effects vanish, the renormalisation factors are in agreement.



One might worry that the large errors appearing in Fig.~\ref{Zss} for 
the $\overline{s}\gamma_\mu c$ renormalisation factors determined here 
would carry forward into our determination of the form factor $f_+$.
However, the vector current matrix element at zero recoil, which contributes the dominant error in $Z_V$, is highly correlated with the vector matrix elements at non-zero recoil.
These correlations cancel in the ratio $\mathcal{V}^0/\mathcal{V}^0(q^2_\text{max})$ appearing when using Eq.~(\ref{PCVClatt}) to 
construct the renormalised current $Z_V \mathcal{V}^0$ appearing in Eqs.~(\ref{fplusextract}) and (\ref{fplusextract2}).
Hence, the uncertainty in the renormalisation factor is not a large contribution to our final uncertainty in the form factors. 

\subsection{Form Factors} \label{subsec:res_ff}
\begin{figure}[t] 
\centering
\includegraphics[width=0.45\textwidth]{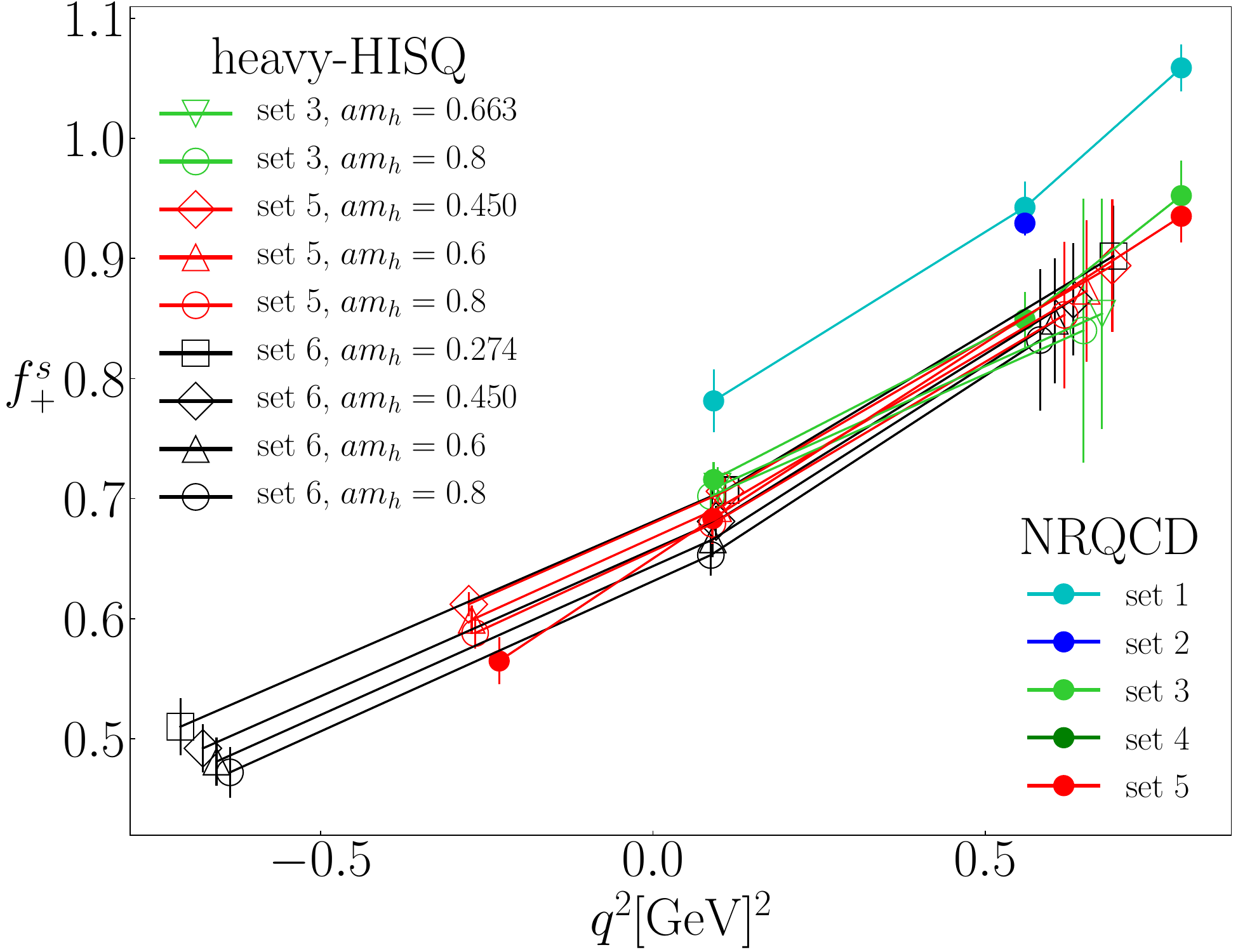}
\caption{$f_+$ form factor data for $B_c^+ \to B_s^0 \overline{\ell} \nu_{\ell}$ from both the NRQCD and heavy-HISQ approaches. 
The NRQCD form factor data is given by filled circles; the heavy-HISQ data, by open circles. Data points on a given set and for a given heavy quark mass are joined by lines to guide eye.}
\label{raw_fp_comp_ctos}
\end{figure}

Fig.~\ref{raw_fp_comp_ctos} provides an example of 
the extracted values for the form factor $f_+$, comparing results from 
the NRQCD and heavy-HISQ spectator calculations. 
The lines on the figures connect data on the same set at a given $am_h$ value 
and are present as a guide only.
The spread of the heavy-HISQ data for different heavy quark masses is 
small, and the NRQCD and heavy-HISQ results are in good agreement 
on the fine lattice.
Discretisation effects are more noticeable for the case of an NRQCD spectator quark, 
especially on the coarsest lattices, sets 1 and 2.
We believe that they result from the $B_c$ meson in the 
calculation since the effects are comparable to those seen in the $B_c$ meson decay 
constant study with NRQCD $b$ quarks in~\cite{Colquhoun:2015oha}.
Data points outside the physical region of momentum transfer are unphysical but nevertheless aid the fit.

\section{Discussion} \label{sec:discussion}
\subsection{$z$ Expansion} \label{fffittingsec}

The four form factors, $f_0$ and $f_+$ for each of the $B_c \to B_s$ and $B_c \to B_d$ processes, at all momenta on all the lattices, are fit simultaneously to a functional form which allows for dependence on the lattice spacing $a$ and bare quark masses.
The fit is carried out using the \textit{lsqfit} package \cite{lsqfit} that implements a least-squares fitting procedure.
As is now standard, we map the semileptonic region $0<q^2<(M_{B_c} - M_{B_{s(d)}})^2$ to a region on the real axis within the unit circle through
\begin{align} \label{littlez}
z(q^2) &= \frac{\sqrt{t_+ - q^2} - \sqrt{t_+ - t_0}}{\sqrt{t_+ - q^2} + \sqrt{t_+ - t_0}},
\end{align}
so that the form factors can be approximated by a truncated power series in $z$.
Here we choose the parameter $t_0$ to be 0 so that the points $q^2 = 0$ and $z = 0$ coincide.
The parameter $t_+$ is in principle the threshold for production of mesons, the lightest being $D$ + $K$, from the $c\overline{s}$ current in the $t$-channel.  
It is convenient here, however, to work with 
$t_+ = (M_{B_c}+M_{B_{s(d)}})^2$, but this gives a very small range for $z$ 
because then $t_+ \gg t_-$.
To correct for this we rescale $z$.

The rescaling factor that we use is $|z(M_{p}^2)|^{-1}$, where $M_{p}$ 
is the mass of the nearest $c\overline{s}$ or $c\overline{d}$ meson pole (we use the same 
mass for both vector and scalar form factors for convenience).
For $B_c \to B_s$ we take $M_p$ as the mass of the vector meson $D_s^{*}$ and for $B_c \to B_d$,
the mass of $D^{*0}$.
Thus, we define 
\begin{align}
\label{eq:zpdef}
z_p (q^2) &= \frac{z(q^2)}{|z(M_{p}^2)|}.
\end{align}
$z_p$ then has a range more commensurate to that for the corresponding 
$D$ decay and the polynomial coefficients in $z_p$ are $\mathcal{O}(1)$. 
Coefficients of the conventional expansion in terms of $z$ can easily be obtained from 
the expansion in $z_p$. Using $z_p$ also avoids introducing large heavy mass dependence 
through the $z$ transform in the heavy-HISQ case, which otherwise would require large $\Lambda_\text{QCD}/M_{H_c}$ coefficients in the heavy-HISQ fit.
Note that in the case of the heavy-HISQ spectator, the $B$-meson masses above in $t_+$ are replaced by the appropriate heavy meson masses at each value of $am_h$ (see Sec.~\ref{subsec:hHISQ_ff_fits}).

\subsection{NRQCD Form Factor Fits} \label{sec:NRQCDFFF}
\begin{figure*}
\centering
\includegraphics[width=1.0\textwidth]{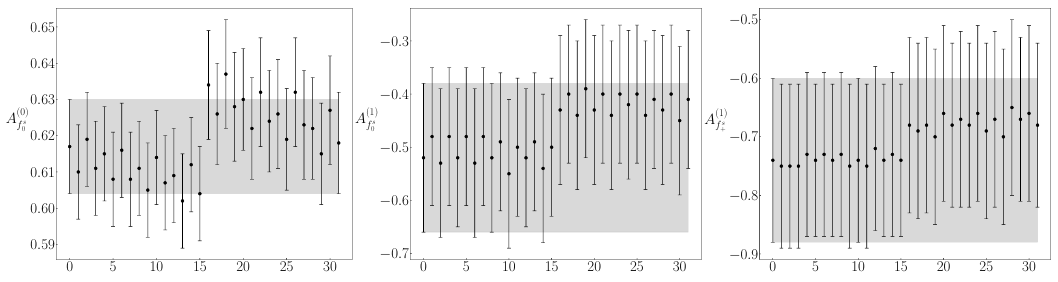}
\caption{$z_p$ expansion coefficients, for the calculation with an NRQCD spectator quark, computed using the variations of correlator fit parameters listed in Table \ref{tab:NRQCDparams} for the $B_c\rightarrow B_s$ form factors. The integer $x$ coordinate of each result is given by $n_\text{var}^{(1)}+2n_\text{var}^{(2)}+4n_\text{var}^{(3)}+8n_\text{var}^{(4)}+16n_\text{var}^{(5)}$ where $n_\text{var}^{(i)} = 0,1$ corresponding to the first and second fits respectively listed in Table \ref{tab:NRQCDparams} of set $i$.}
\label{stabplotNRQCD1}
\end{figure*}
\begin{figure*}
\centering
\includegraphics[width=1.0\textwidth]{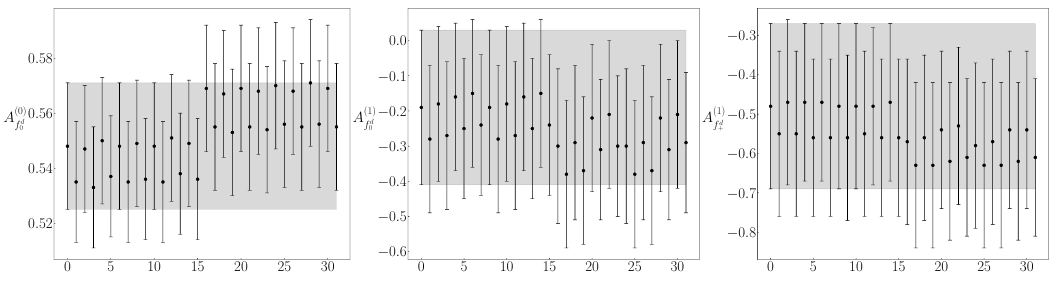}
\caption{$z_p$ expansion coefficients, for the calculation with an NRQCD spectator quark, computed using the variations of correlator fit parameters listed in Table \ref{tab:NRQCDparams} for the $B_c\rightarrow B_d$ form factors. The $x$ coordinate is the same as that in Fig.~\ref{stabplotNRQCD1}.}
\label{stabplotNRQCD2}
\end{figure*}
The form factor results from the calculation with NRQCD spectator 
quark are fit to
\begin{align} \label{fffitform}
f(q^2) = & \hspace{1mm} P(q^2) \sum_{n=0}^{N} b^{(n)} z_p^n .
\end{align}
Here, the dominant pole structure is represented by a factor $P(q^2)$ given by $(1-q^2/M^2_{\text{res}})^{-1}$ 
with $M_{\text{res}}$ the mass of the relevant $c\overline{s}$ or $c\overline{d}$ meson 
(the vector meson for $f_+$ and the scalar for $f_0$). We take the values of 
$M_{\text{res}}$ from current experiment~\cite{PDG}:
$M_{D_s^*}=2.112$ GeV, $M_{D_{s0}^*}=2.317$ GeV, $M_{D^*}=2.01027$ GeV, and $M_{D_0^*}=2.300$ GeV. 
We do not include uncertainties in these values since $P(q^2)$ is a purely fixed 
factor designed to remove much of the $q^2$-dependence from the form factors. 
For our lattice results uncertainties enter $P(q^2)$ from the uncertainty in our 
determination of $q^2$ in physical units, including that from the determination 
of the lattice spacing. 

$P(q^2)$ multiplies a polynomial in $z_p$, and the polynomial coefficients are
\begin{align} \label{fffitform2}
b^{(n)}= & \hspace{1mm} A^{(n)} \Big\{ 1+ B^{(n)} (am_c/\pi)^2 + C^{(n)} (am_c/\pi)^4 \nonumber \\ 
+ & \kappa_{1}^{(n)} \frac{\delta m_l^{\text{sea}}}{10 m_s^{\text{tuned}}} + \kappa_{2}^{(n)} \frac{\delta m_s^{\text{sea}}}{10 m_s^{\text{tuned}}} + \kappa_{3}^{(n)} \frac{\delta m_c^{\text{sea}}}{m_c^{\text{tuned}}} \nonumber \\
+ & \kappa^{(n)}_{4} \frac{\delta m_{s}^{\text{val}}}{10 m_s^{\text{tuned}}} + \kappa^{(n)}_5 \frac{\delta m_c^{\text{val}}}{m_c^{\text{tuned}}} + \kappa^{(n)}_6 \frac{\delta m_b^{\text{val}}}{m_b^{\text{tuned}}} \Big\}. \nonumber \\
\end{align}

The parameters $\kappa^{(n)}_j$ allow for errors associated with mistunings of both sea and valence quark masses.
The term accounting for mistuning of valence strange quarks is included only for the $B_c \to B_s$ transition.
The tuned masses $m_s^{\text{tuned}}$ and $m_c^{\text{tuned}}$ are the valence quark masses that yield physical $\eta_s$ and $\eta_c$ meson masses respectively in the sea of 2+1+1 flavours of sea quark.
Values for $m_s^{\text{tuned}}$ and $m_c^{\text{tuned}}$ were obtained from~\cite{{Chakraborty:2014aca}}.
%
Also, $m_l^{\text{tuned}}$ is fixed by multiplying $m_s^{\text{tuned}}$ by the physical ratio
\begin{align} 
\frac{m_l}{m_s} = & \hspace{1mm}  \frac{1}{27.18(10)} 
\end{align}
obtained from \cite{Bazavov:2017lyh}.
For the $b$ quark, we take tuned values\footnote{To ensure consistency, we convert values from~\cite{Dowdall:2011wh} in lattice units to physical units by using the lattice spacing determined in~\cite{Dowdall:2011wh} from the $\Upsilon(2S-1S)$ splitting.} of the quark mass from Table XII in~\cite{Dowdall:2011wh}.

For each of the sea and valence quark flavours, $\delta m^{\text{sea}}$ and $\delta m^{\text{val}}$ are given by
\begin{align} 
\delta m^{\text{sea}} = & \hspace{1mm}  m^{\text{sea}} - m^{\text{tuned}} \nonumber \\
\delta m^{\text{val}} = & \hspace{1mm}  m^{\text{val}} - m^{\text{tuned}},
\end{align}
giving estimates of the extent that the quark masses deviate from the ideal choices in which appropriate meson masses are exactly reproduced.

For prior values on the parameters in Eq.~(\ref{fffitform2}), we use $0(1)$ for $A^{(n)}$, $B^{(n)}$ and $C^{(n)}$, and $0.0(5)$ for $\kappa_{(j)}$.
The power series in Eq.~(\ref{fffitform}) is truncated to include up to the $z_p^3$ term.
Fits without a pole, i.e. $P(q^2) = 1$, yield no statistically significant discrepancies.
This is not surprising since the poles are far away from the physical region of $q^2$, and so the pole effect on the form factor can be reasonably absorbed into the polynomial.
Finally, the kinematic relation
\begin{align}
f_0 (0) = f_+ (0)
\end{align}
is imposed on the fit as a constraint (we have tested that removing this constraint makes very little 
difference to the fit in fact and $f_+(0)-f_0(0)$ is zero to well within 1$\sigma$.). 

Constraints on $b^{(n)}$ from unitarity, as in the Bourrely-Caprini-Lellouch (BCL)~\cite{Bourrely:2008za} 
and Boyd-Grinstein-Lebed (BGL)~\cite{Boyd:1995sq} expansions, are  unnecessary here since the full range of physical momentum transfer can be reached and so extrapolation in $q^2$, which may benefit in accuracy from imposing these constraints, is not required.
Hence, more complicated fit forms that impose additional physical constraints are not expected to be appreciably advantageous.

In Figs.~\ref{stabplotNRQCD1} and \ref{stabplotNRQCD2}, we demonstrate that the form factors in the physical-continuum limit are insensitive to the choice of the parameters in the fits of the correlators.
As can be seen in the figures, the coefficients in the fits of the form 
factors are stable, within their uncertainties, as the correlator fits on different sets are varied.

\begin{figure}
\centering
\includegraphics[width=0.45\textwidth]{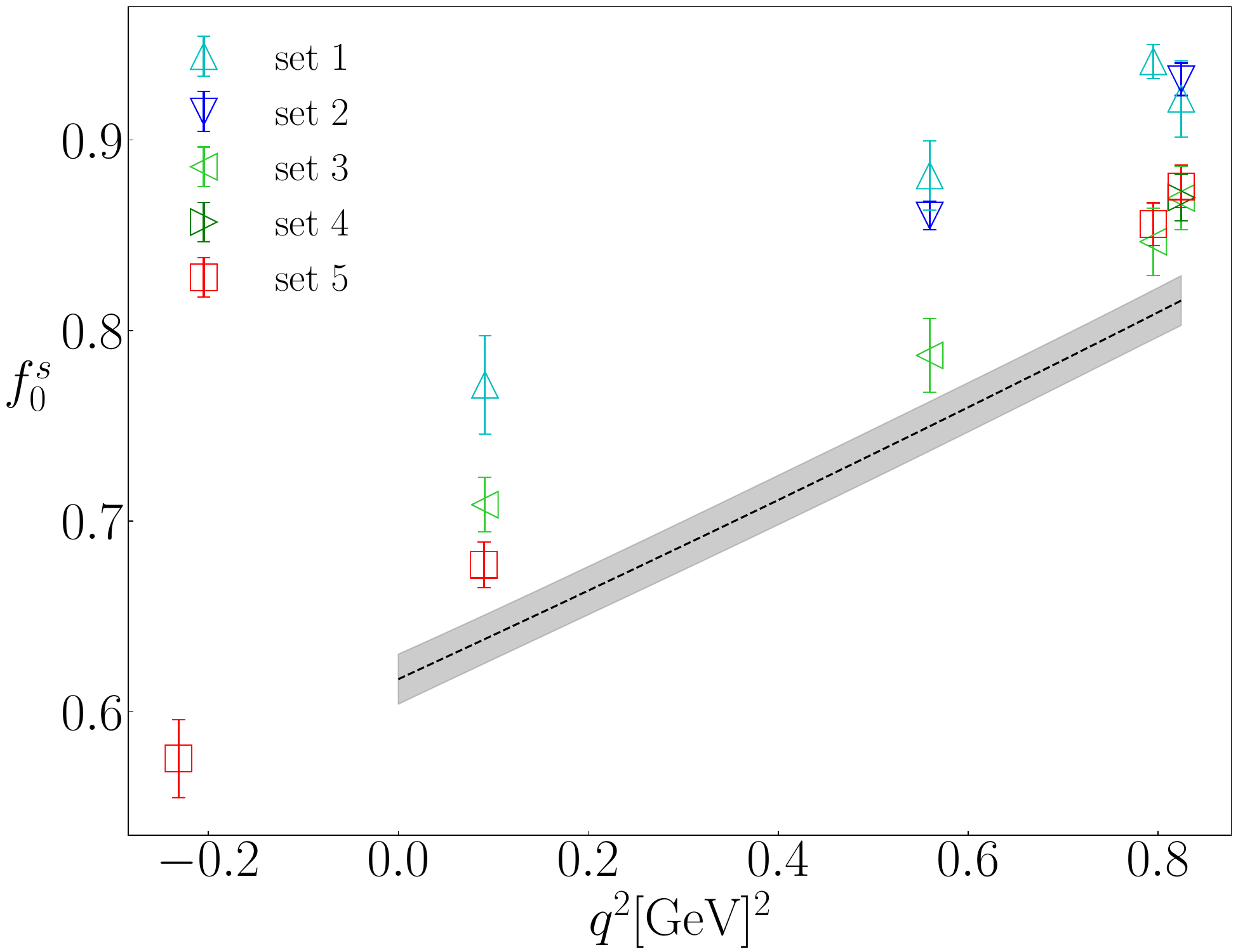}
\caption{Lattice results and fitted $f_0$ form factor data for $B_c^+ \to B_s^0 \overline{\ell} \nu_{\ell}$ with an NRQCD $b$ quark. 
The grey band shows the fitted form factor tuned to the limit of vanishing lattice spacing and physical quark masses.}
\label{0s}
\end{figure}

\begin{figure}
\centering
\includegraphics[width=0.45\textwidth]{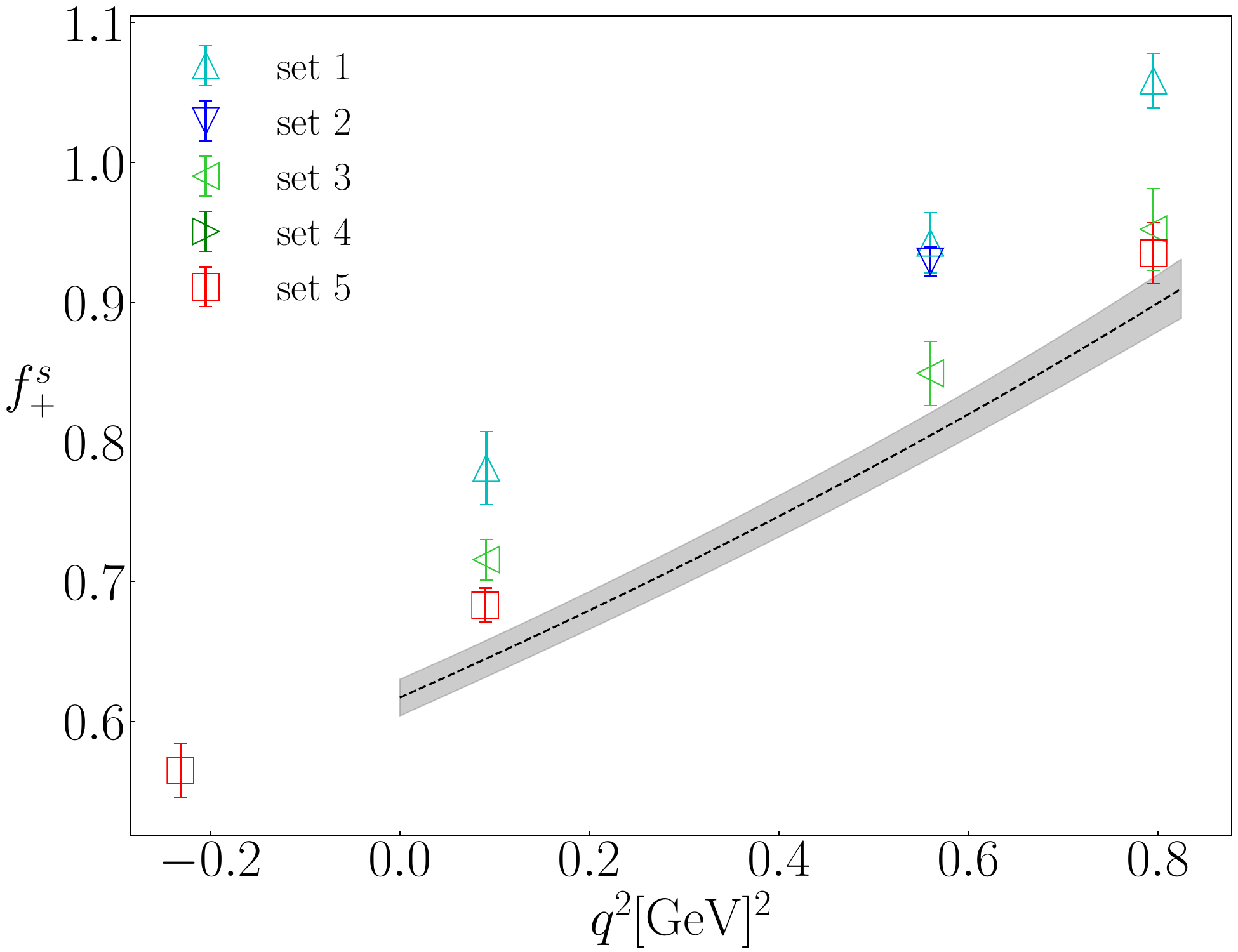} 
\caption{Lattice results and fitted $f_+$ form factor data for $B_c^+ \to B_s^0 \overline{\ell} \nu_{\ell}$ with an NRQCD $b$ quark.
The grey band shows the fitted form factor tuned to the limit of vanishing lattice spacing and physical quark masses.
}
\label{ps}
\end{figure}

\begin{figure} 
\centering
\includegraphics[width=0.45\textwidth]{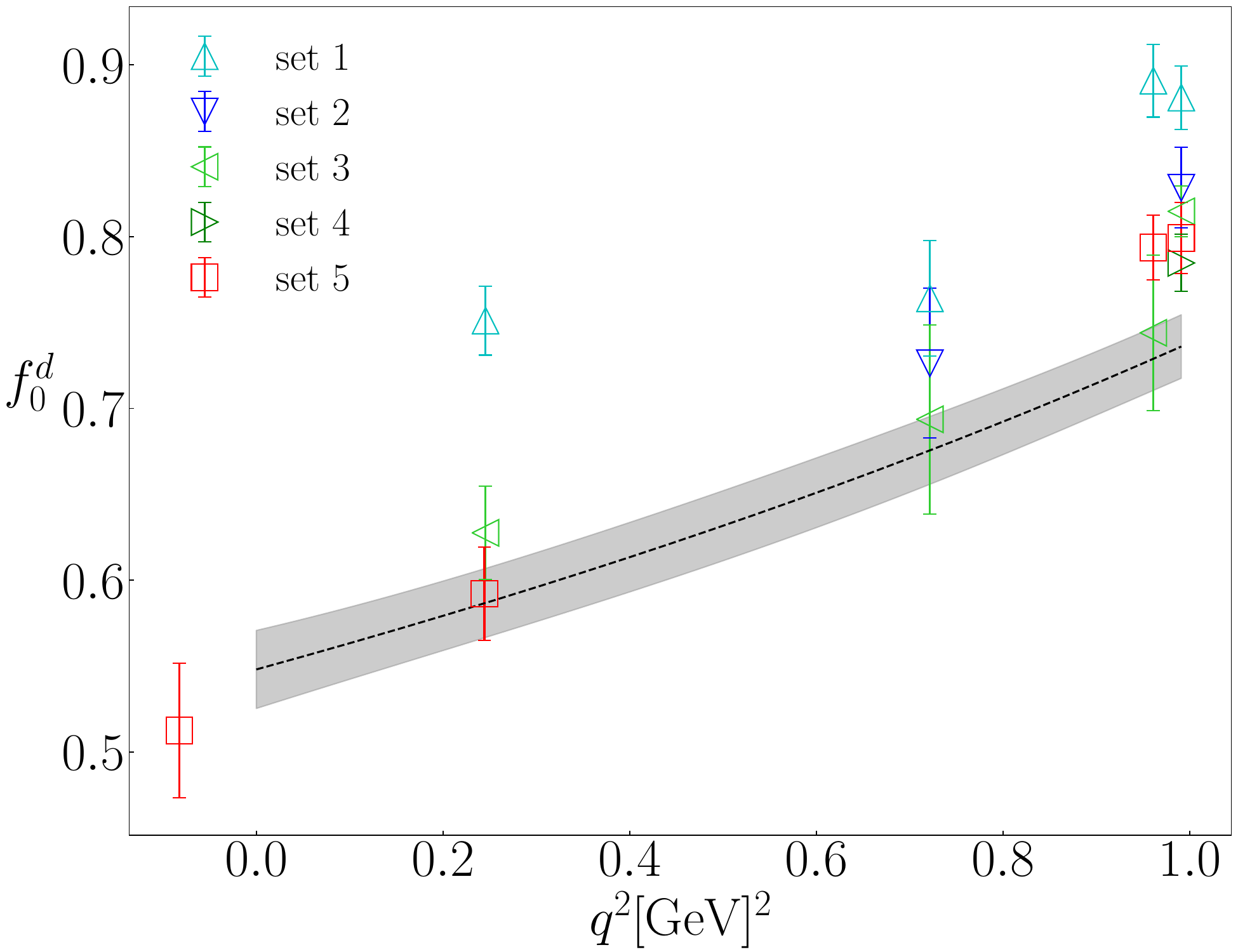}
\caption{Lattice results and fitted $f_0$ form factor data for $B_c^+ \to B^0 \overline{\ell} \nu_{\ell}$ with an NRQCD $b$ quark.
The grey band shows the fitted form factor tuned to the limit of vanishing lattice spacing and physical quark masses.
}
\label{0d}
\end{figure}

\begin{figure}
\centering
\includegraphics[width=0.45\textwidth]{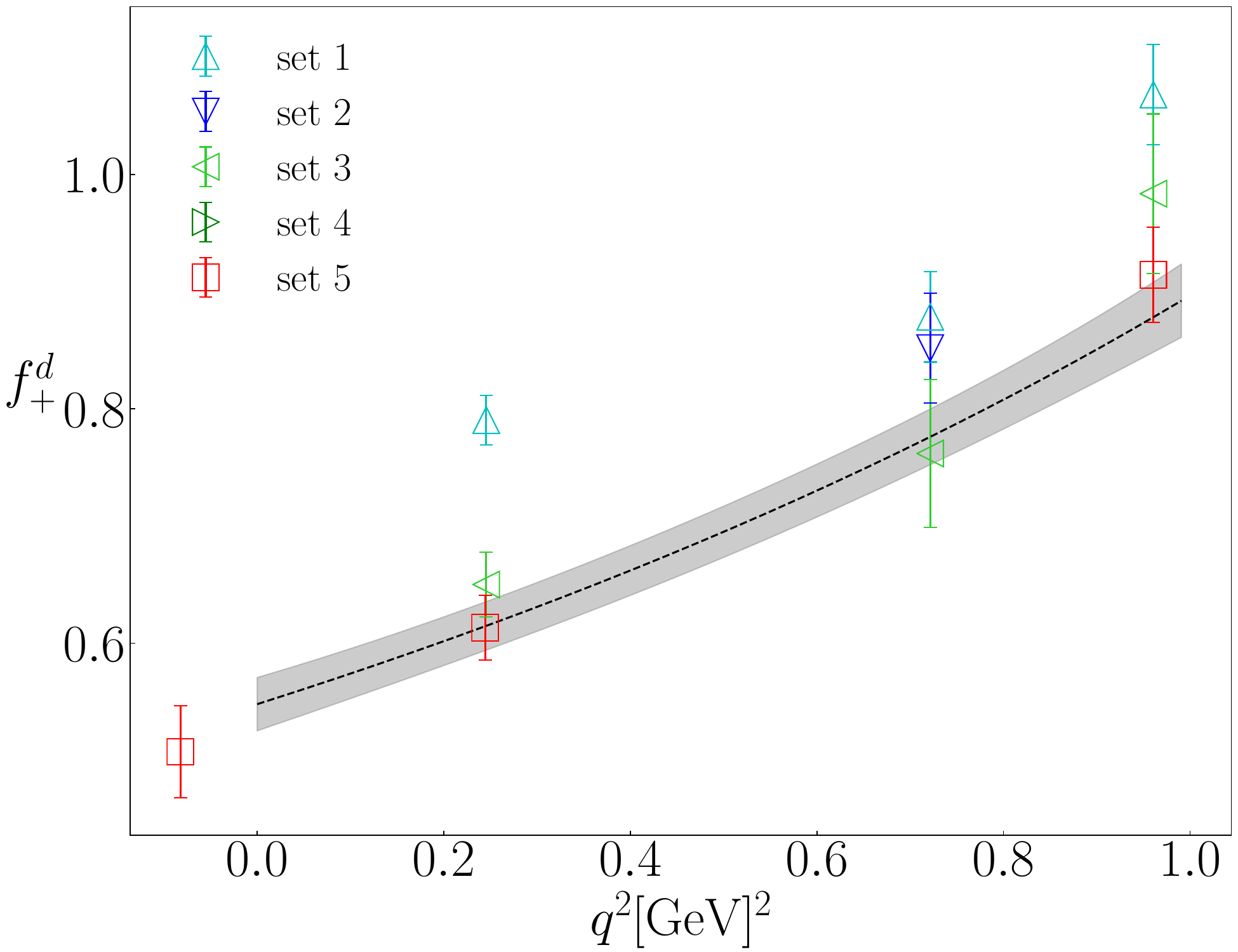} 
\caption{Fitted $f_+$ form factor data for $B_c^+ \to B^0 \overline{\ell} \nu_{\ell}$ with an NRQCD $b$ quark.
The grey band shows the fitted form factor tuned to the limit of vanishing lattice spacing and physical quark masses.
}
\label{pd}
\end{figure}

The fitted form factors from the NRQCD spectator case exhibit errors no greater than $4\%$ across the entire physical range of $q^2$ when tuned to the physical-continuum limit.
Figs.~\ref{0s}, \ref{ps}, \ref{0d} and \ref{pd} show the results on all the lattices 
along with the fitted function for the form factors in the physical-continuum limit.



\begin{table}
\centering
\caption{A selection of $f_0$ and $f_+$ fit parameters from our fit to 
Eq.~(\ref{fffitform2}) with an NRQCD-$b$ quark, demonstrating the leading order momentum and lattice spacing dependence.
Note that the discretisation effects in the $f_0$ and $f_+$ fits are allowed to vary independently 
of each other with separate $B^{(0)}$ parameters. In practice, as the Table shows, the fit 
returns very similar values.  }
\begin{tabular}{c c c c c} 
 \hline\hline
 & $f_0^s$ & $f_0^d$ & $f_+^s$ & $f_+^d$ \\ [0.1ex] 
\hline
$A^{(0)}$ & 0.617(13) & 0.548(23) & 0.617(13) & 0.548(23) \\
$A^{(1)}$ & -0.52(14) & -0.19(22) & -0.74(14) & -0.48(21) \\
$A^{(2)}$ & -0.63(63) & 0.05(74) & -0.29(72) & 0.12(77) \\
$B^{(0)}$ & 1.44(38) & 1.45(46) & 1.44(38) & 1.45(46) \\
 \hline\hline
\end{tabular}
\label{f0param}
\end{table}

The $z_p^0$ and $z_p^1$ behaviour of the form factors is well resolved by our fit 
to Eq.~(\ref{fffitform2}), as well as the $(a m_c/\pi)^2z_p^0$ discretisation effect.
Table~\ref{f0param} summarises the corresponding parameters from the fit.
After fitting, other parameters show errors comparable to the width of their prior and are consistent with 0.
In particular, quark mass mistuning coefficients simply return their prior value.

\subsection{Heavy-HISQ Form Factor Fits} \label{subsec:hHISQ_ff_fits}
We take a similar approach to fitting the form factor results for the case 
of a heavy-HISQ spectator. Now we have results at multiple heavy-quark masses 
and the conversion from $q^2$ to $z$-space (Eq.~(\ref{littlez}))
uses the values of $M_{H_c}$ and $M_{H_s}$ or $M_{H_d}$, as appropriate, from 
our calculation.  We then rescale $z$ at each $m_h$ as described in 
Sec.~\ref{fffittingsec} (Eq.~(\ref{eq:zpdef})). This rescaling gives 
a similar $z$-range for each $m_h$ and avoids introducing spurious dependence 
on $m_h$ that comes simply from the $z$-transform.  
I
The heavy-HISQ results are then fit to a form that is a product of $P(q^2)$ and 
a polynomial in $z_p$ as for the NRQCD case. We now require a fit form for the polynomial 
coefficients that accounts 
for $(am_h)^{2n}$ discretisation effects as well as physical dependence on $m_h$, however. Motivated by heavy quark effective theory (HQET) we express this physical heavy mass dependence as a power series in $\Lambda_\text{QCD}/M_{H_c}$.
The form factor data from the heavy-HISQ approach is fit to
\begin{align}
f (q^2) = P(q^2)&\sum_{n,i,j,k = 0}^{3} A^{(n)}_{ijk} z_p^n \nonumber\\
&\times\left(\frac{am_c}{\pi}\right)^{2i} \left(\frac{am_h}{\pi}\right)^{2j} \Delta_{H_c}^{(k)} \mathcal{N}_\text{mis}^{(n)}, \label{hhisqffff}
\end{align}
where, for $k=0$, $\Delta_{M}^{(k)}=1$ and, for $k \neq 0$,
\begin{equation}
\Delta_{H_c}^{(k)} =\left(\frac{\Lambda_\text{QCD}}{M_{H_c}}\right)^k- \left(\frac{\Lambda_\text{QCD}}{M_{B_c}}\right)^k 
\end{equation}
where we take $\Lambda_{QCD} = 500\text{MeV}$. The mistuning terms are given by
\begin{align}
\mathcal{N}_\text{mis}^{(n)}& = 1 + \frac{\delta m_c^{\text{val}}}{m_c^\text{tuned}}a_n + \frac{\delta m_c^{\text{sea}}}{m_c^\text{tuned}}b_n \nonumber\\
+ \frac{\delta m_s^{\text{val}}}{10 m_s^\text{tuned}}&c_n + \frac{\delta m_s^{\text{sea}}}{10m_s^\text{tuned}}d_n + \frac{\delta{m_l}}{10m_s^\text{tuned}}e_n,
\end{align}
where we only include the term proportional to $\delta m_s^\text{val}$ for 
the $B_c\rightarrow B_s$ case. $P(q^2)$,  $\delta{m}$ and the tuned masses 
have the same definitions as in the NRQCD case (Sec.~\ref{sec:NRQCDFFF}). 
In the physical continuum limit, this form collapses to $P(q^2)\sum_n z_p^n A^{(n)}_{000}$. Again we apply the constraint 
$f_0 (0) = f_+ (0)$ in the continuum limit (by fixing $A_{000}^{(0)}$ to be the same 
in the two cases).

\begin{figure}
\centering
\includegraphics[width=0.45\textwidth]{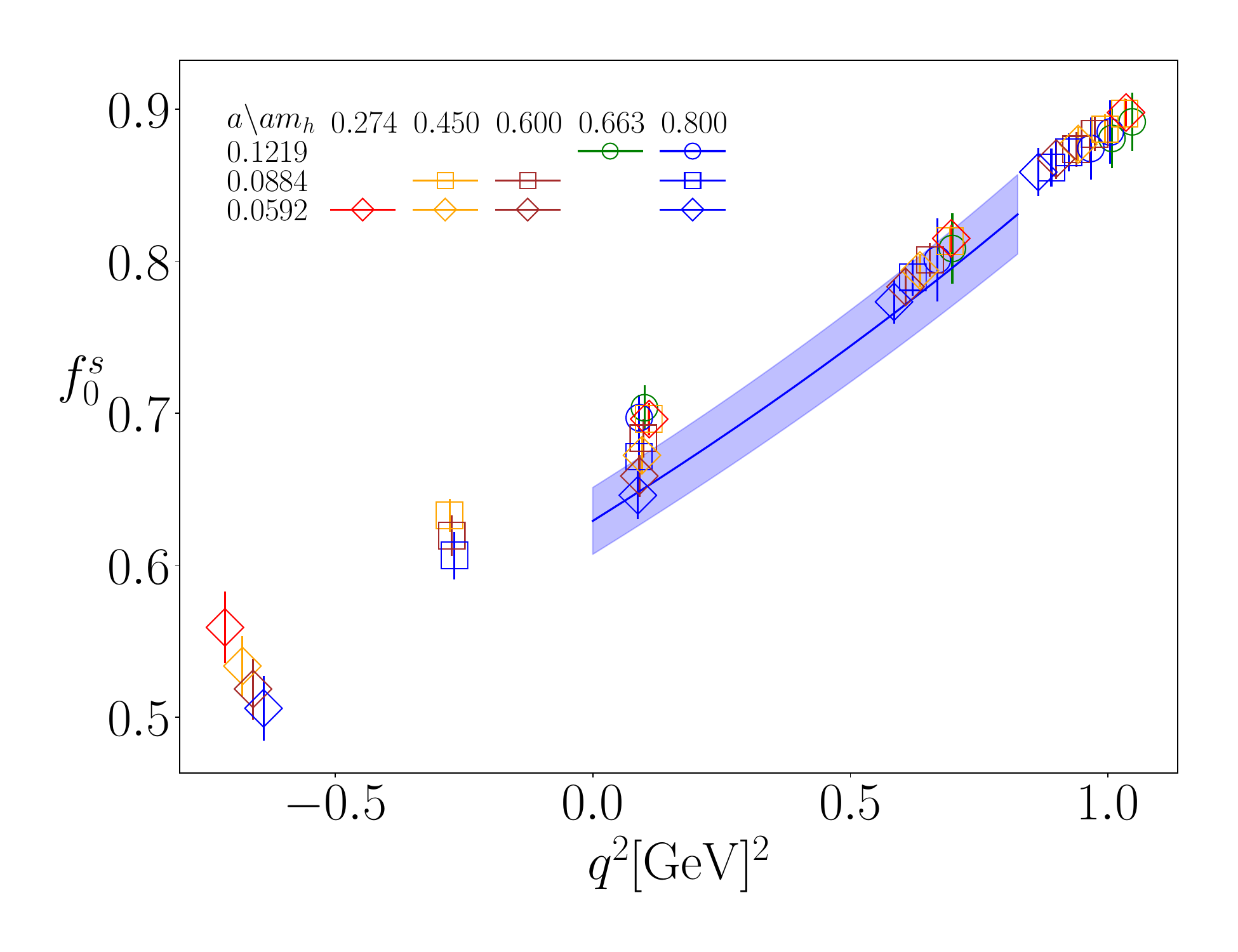}
\caption{Heavy-HISQ form factor results for $f^s_0$ together with the fitted curve 
at the physical point with its error band.}
\label{HHzpfitplotsfs0}
\end{figure}

\begin{figure}
\centering
\includegraphics[width=0.45\textwidth]{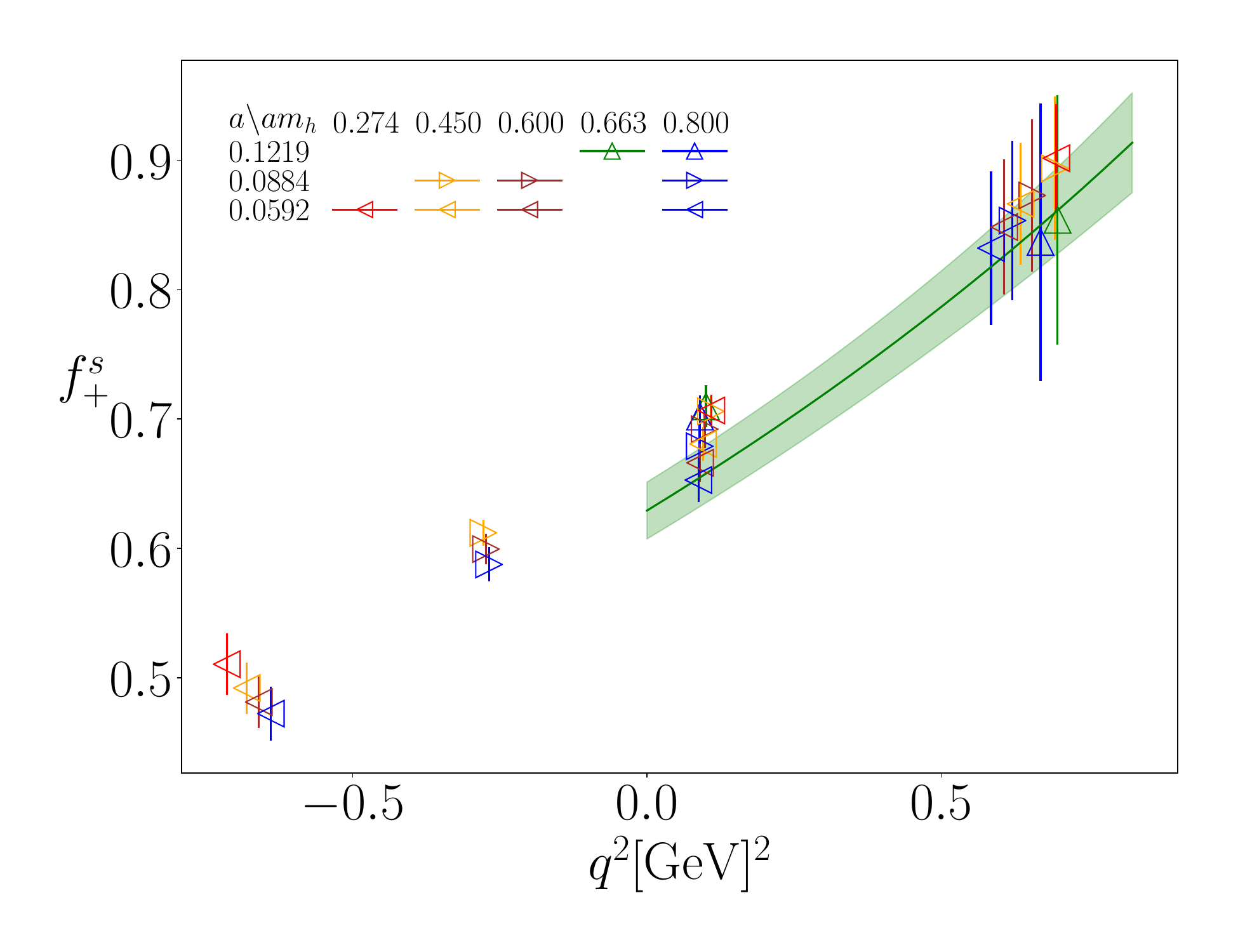}
\caption{Heavy-HISQ form factor results for $f_+^s$ together with the fitted curve 
at the physical point with its error band. }
\label{HHzpfitplotsfsp}
\end{figure}

\begin{figure}
\centering
\includegraphics[width=0.45\textwidth]{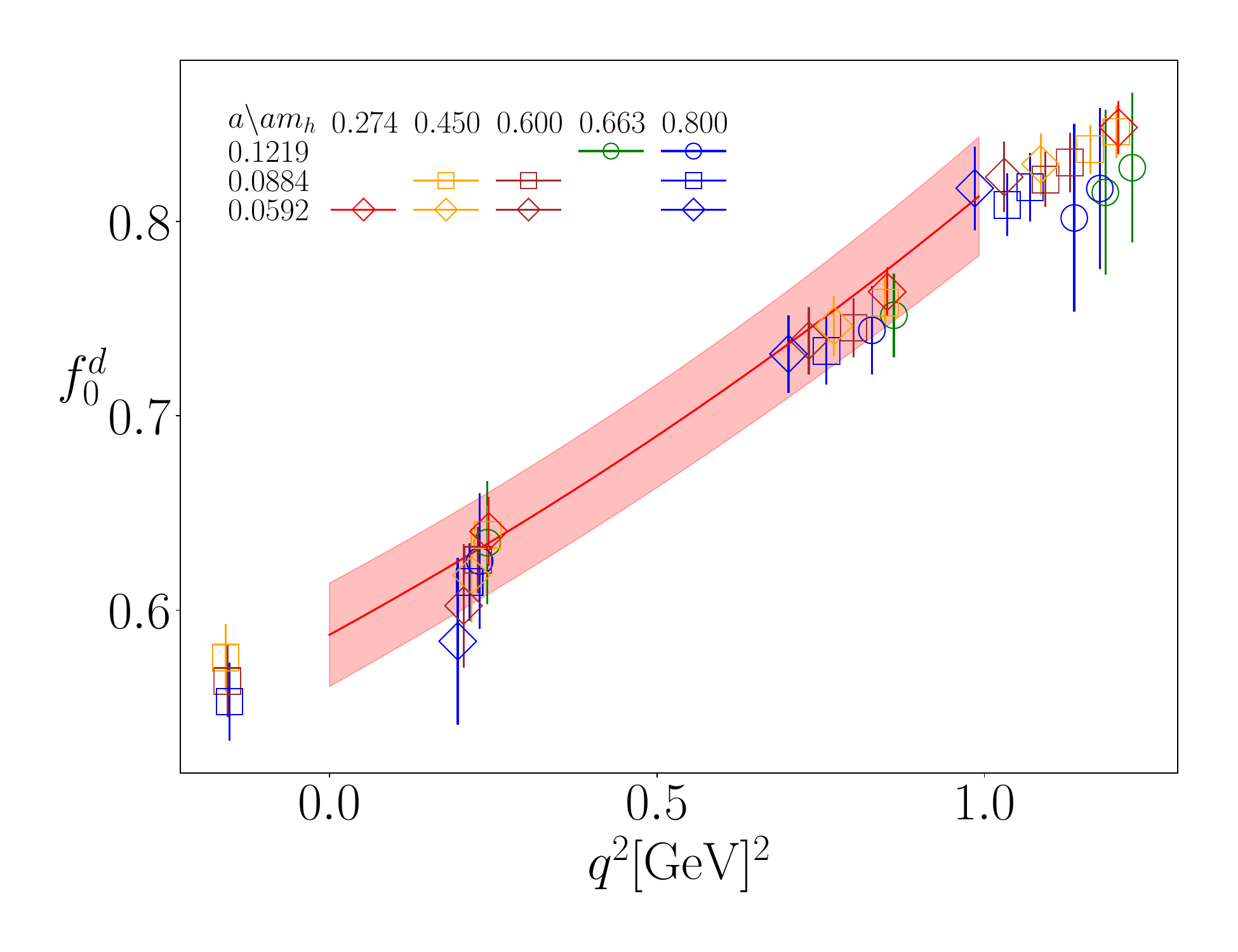}
\caption{Heavy-HISQ form factor results for $f^d_0$ together with the fitted curve 
at the physical point with its error band.}
\label{HHzpfitplotsfd0}
\end{figure}

\begin{figure}
\centering
\includegraphics[width=0.45\textwidth]{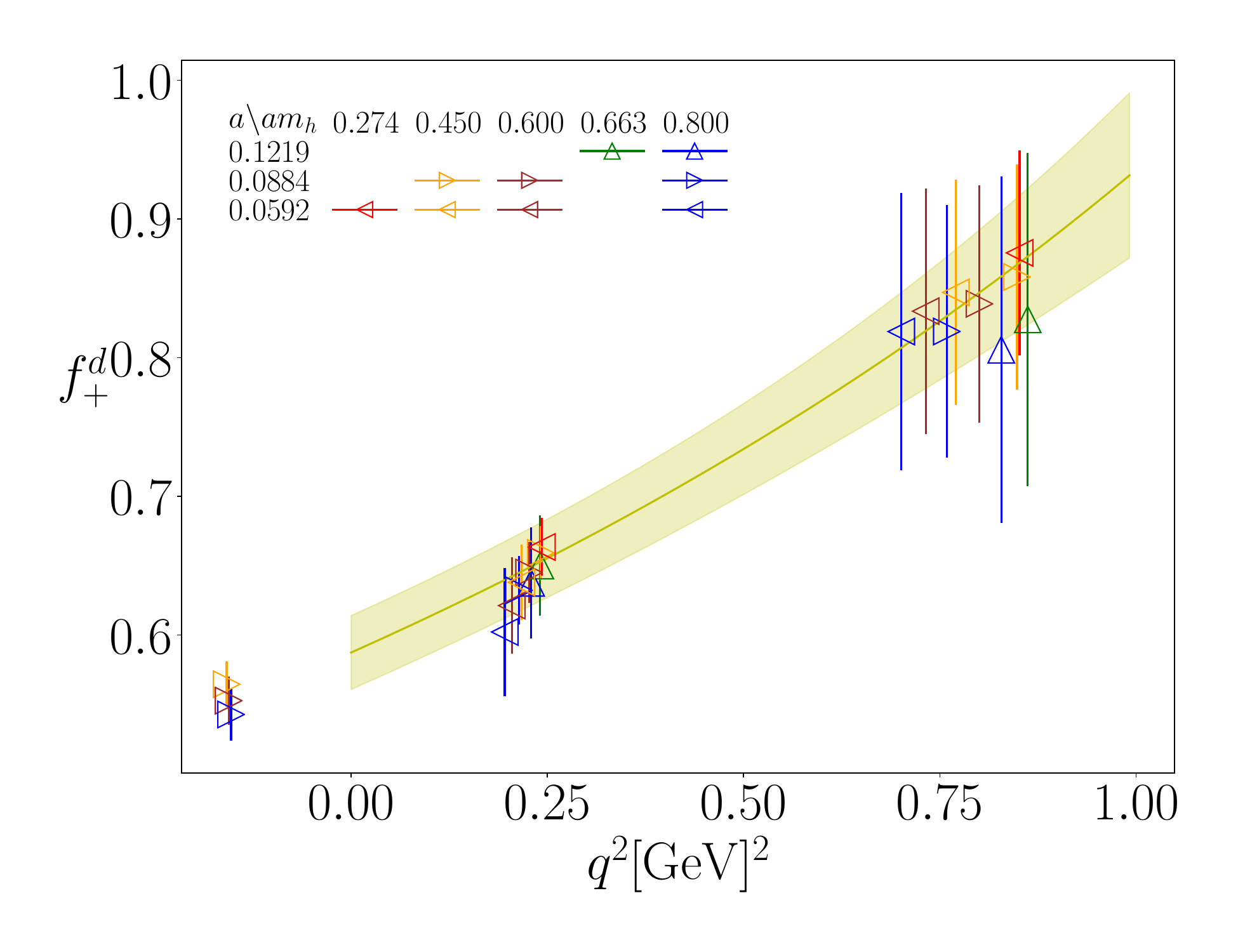}
\caption{Heavy-HISQ form factor results for $f_+^d$ together with the fitted curve 
at the physical point with its error band. }
\label{HHzpfitplotsfdp}
\end{figure}

Results for the extrapolated form factors are given in Figs.~\ref{HHzpfitplotsfs0}, \ref{HHzpfitplotsfsp}, \ref{HHzpfitplotsfd0} and \ref{HHzpfitplotsfdp} together with the corresponding lattice data. For the $B_c\rightarrow B_s$ case $a_n$ and $c_n$ take prior values $0(1)$ and $b_n$, $d_n$ and $e_n$ take prior values $0(0.3)$ to reflect the fact that they enter through loop effects. In the $B_c\rightarrow B_d$ case we take prior values of $0(1)$ for $a_n$ and $e_n$ and $0(0.3)$ for $b_n$ and $d_n$. In both cases we take prior values of $0(1)$ for $A_{ijk}^n$ except for when $i=1$ or $j=1$ where we use a prior values of $0(0.3)$ to account for the HISQ one loop improvement. 

\begin{figure*}
\centering
\includegraphics[width=1.0\textwidth]{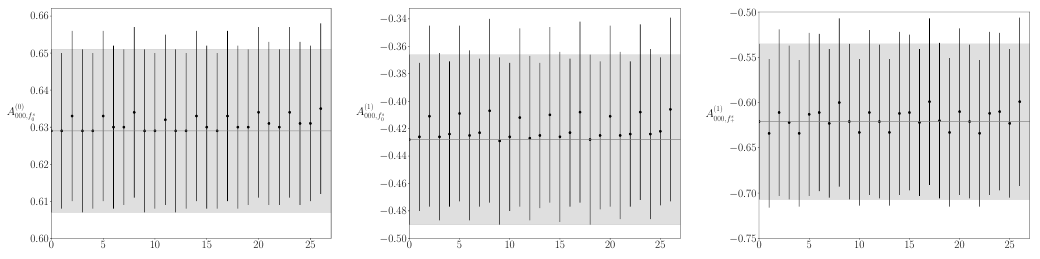}
\caption{$z_p$ expansion coefficients, for the calculation with a HISQ spectator quark, computed using the variations of correlator fit parameters listed in Table \ref{HHparams} for the $B_c\rightarrow B_s$ form factors. The integer $x$ coordinate of each result is given by $n_\text{var}^{(3)}+3n_\text{var}^{(5)}+9n_\text{var}^{(6)}$ where $n_\text{var}^{(i)} = 0,1,2$ corresponding to original, first and second variations respectively listed in Table \ref{HHparams} of set $i$.}
\label{stabplothh1}
\end{figure*}

\begin{figure*}
\centering
\includegraphics[width=1.0\textwidth]{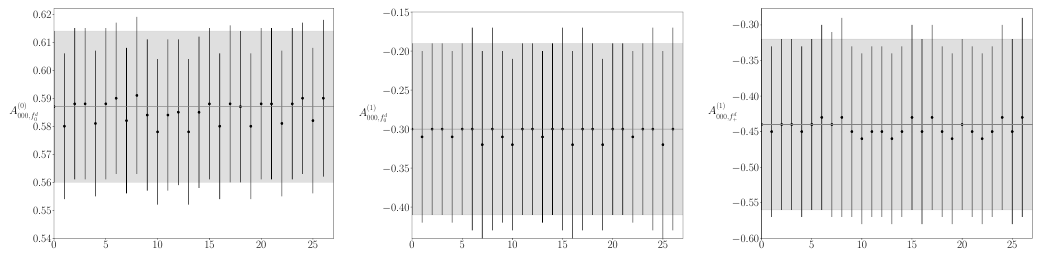}
\caption{$z_p$ expansion coefficients, for the calculation with a HISQ spectator quark, computed using the variations of correlator fit parameters listed in Table \ref{HHparams} for the $B_c\rightarrow B_d$ form factors. The $x$ coordinate is the same as that in Fig.~\ref{stabplothh1}.}
\label{stabplothh2}
\end{figure*}

As in the case of an NRQCD spectator quark, we present coefficients of the form factors fits from many different fits of the correlator data.
Figs.~\ref{stabplothh1} and \ref{stabplothh2}, show that the coefficients are insensitive to the choice of the parameters in the fits of the correlators.

\subsection{Chained Fit} \label{sec:ff_chained_fit}
\begin{figure}
\centering
\includegraphics[width=0.45\textwidth]{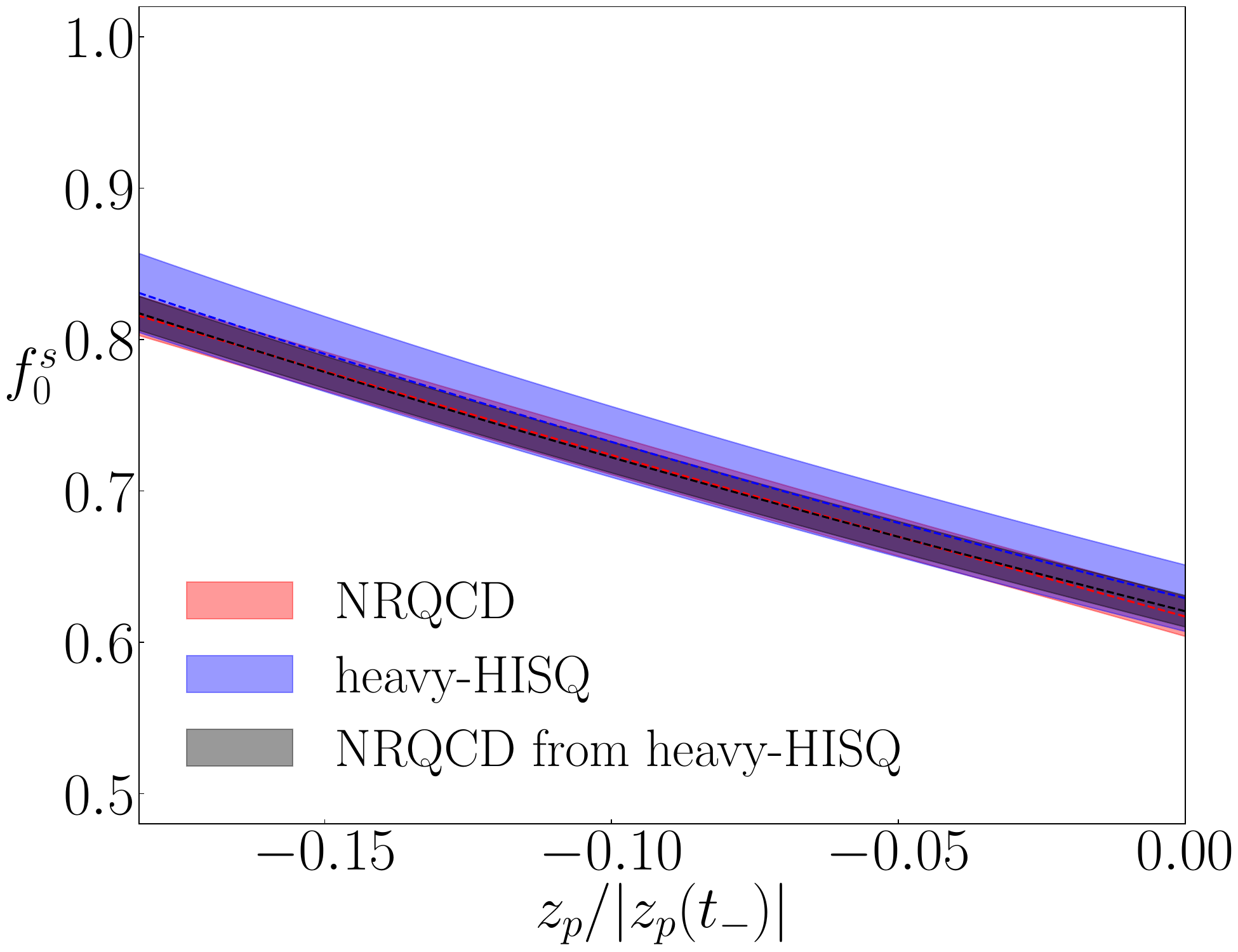}
\caption{Fits of $f_0$ for $B_c^+ \to B_s^0 \overline{\ell} \nu_{\ell}$ tuned to the physical-continuum limit. The form factor is plotted against $z_p/|z_p(t_-)|$.}
\label{f0_physcont_comp_strange}
\end{figure}

\begin{figure}
\centering
\includegraphics[width=0.45\textwidth]{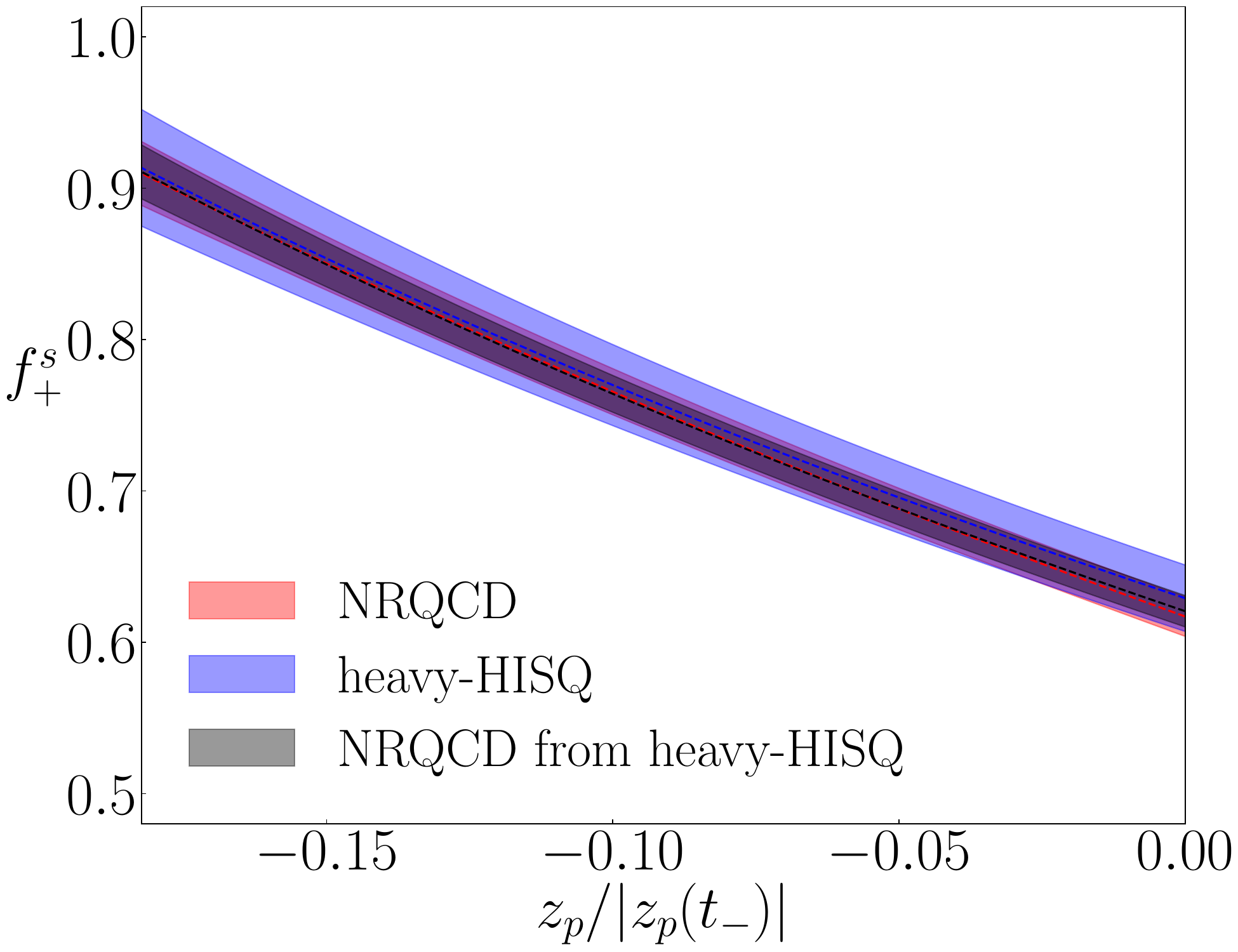}
\caption{Fits of $f_+$ for $B_c^+ \to B_s^0 \overline{\ell} \nu_{\ell}$ tuned to the physical-continuum limit.}
\label{fplus_physcont_comp_strange}
\end{figure}

\begin{figure}
\centering
\includegraphics[width=0.45\textwidth]{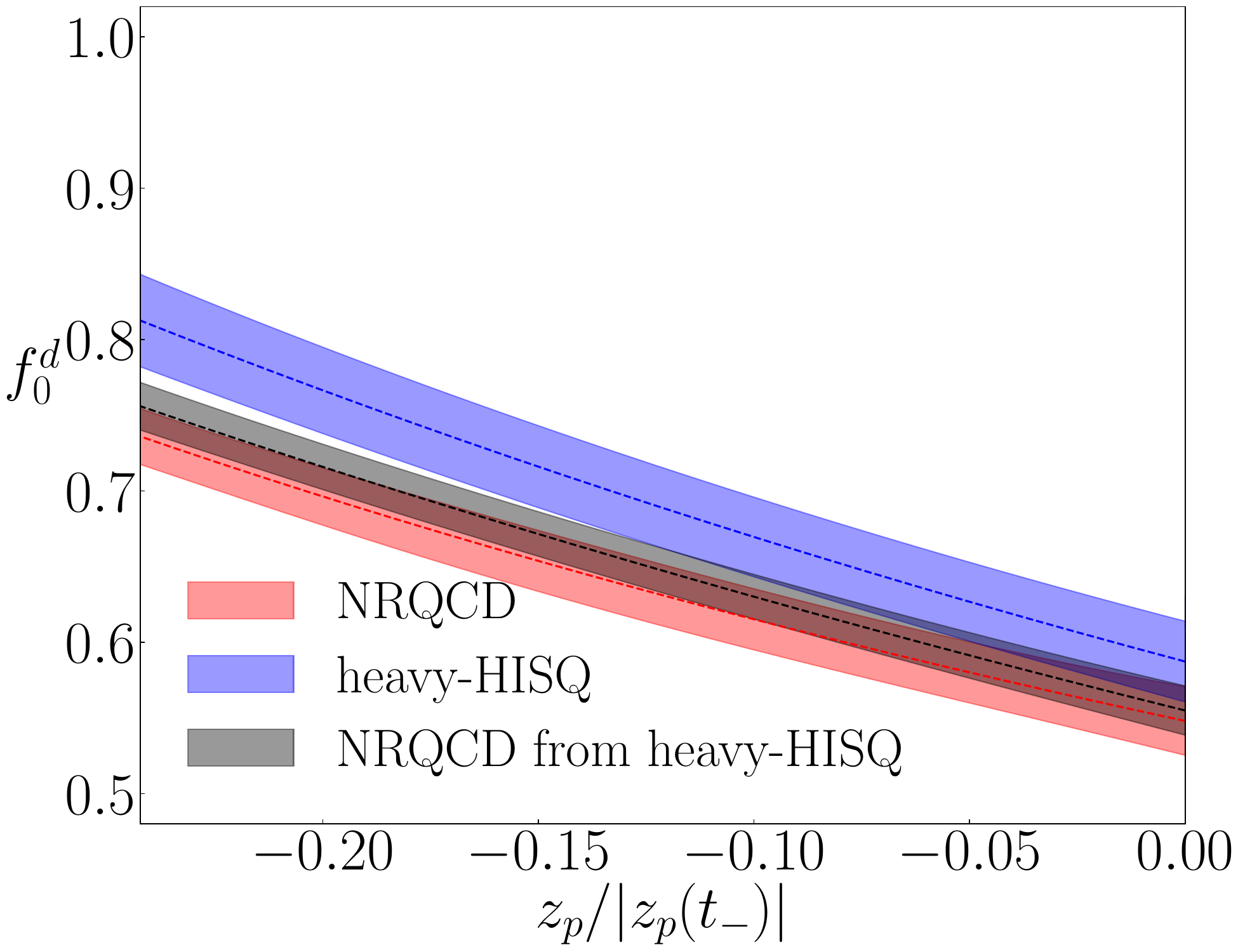}
\caption{Fits of $f_0$ for $B_c^+ \to B^0 \overline{\ell} \nu_{\ell}$ tuned to the physical-continuum limit.}
\label{f0_physcont_comp_down}
\end{figure}

\begin{figure}
\centering
\includegraphics[width=0.45\textwidth]{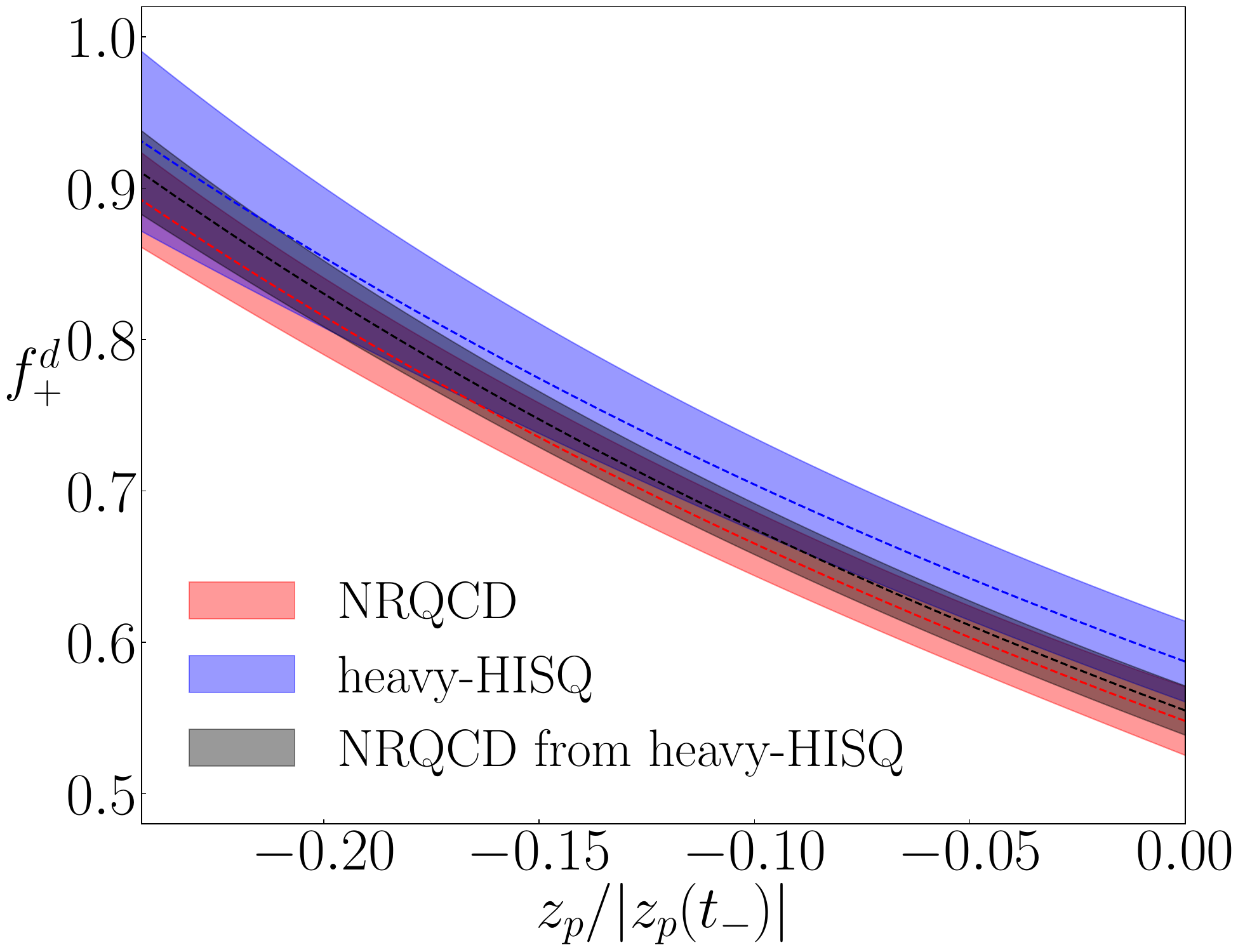}
\caption{Fits of $f_+$ for $B_c^+ \to B^0 \overline{\ell} \nu_{\ell}$ tuned to the physical-continuum limit.}
\label{fplus_physcont_comp_down}
\end{figure}

The form factor functions tuned to the physical-continuum limit from NRQCD and heavy-HISQ  are 
compared in Figs.~\ref{f0_physcont_comp_strange}, \ref{fplus_physcont_comp_strange}, 
\ref{f0_physcont_comp_down} and \ref{fplus_physcont_comp_down} in $z$-space. 
There is good agreement across the entire physical range of $z$, with particularly good 
agreement for the more accurate $B_c \rightarrow B_s$ case.

Whilst the fit forms for the form factors from NRQCD and heavy-HISQ at Eqs.~(\ref{fffitform}) and~(\ref{hhisqffff}) differ in appearance, they both allow for effects of discretisation and mistuning of the quark masses.
In the continuum limit with physical masses, the two forms collapse such that the parameters $A^{(n)}$ from Eq.~(\ref{fffitform2}) and $A_{000}^{(n)}$ from Eq.~(\ref{hhisqffff}) coincide.
Plotted among the functions from the heavy-HISQ and NRQCD calculations is a function arising from a `chained' fit where the $A_{000}^{(n)}$ from the heavy-HISQ fit were used as prior distributions for the $A^{(n)}$ in the form factor fit forms in the NRQCD study.
We label this fit \emph{NRQCD from heavy-HISQ} in Figs.~\ref{f0_physcont_comp_strange}, \ref{fplus_physcont_comp_strange}, \ref{f0_physcont_comp_down} and \ref{fplus_physcont_comp_down}.
As with the separate fits for each case of spectator quark, the form factors for $B_c \to B_s$ and $B_c \to B_d$ are fit simultaneously.
This chained fit has $\chi^2/\text{d.o.f.} = 1.1$ and is consistent with both the separate fits.
We make our final predictions for the decay rates and values for $\Gamma |V|^2$ 
using the chained fit.


We include the coefficients $A_{0,+}^{(n)}$ from the chained fit in the ancillary
\emph{json} file \texttt{BcBsd\_ff\_updated.json}.
To assist those who wish to import the form factors from these coefficients, we provide values that the form factors should take for both $q^2 = 0$ and $q^2_{\mathrm{max}}$ in Table~\ref{tab:ff_extrema}.
\begin{table}
	\centering
	\caption{The form factors from the chained fit evaluated at $q^2 = 0$ and $q^2_{\mathrm{max}}$.}
	\begin{tabular}{c c c | c c} 
		\hline\hline
		$q^2 \; [\mathrm{GeV}]^2$ & $f_0^s$& $f_+^s$& $f_0^d$& $f_+^d$  \\
		\hline
		$0$ & $0.621(10)$ & $0.621(10)$ & $0.555(16)$ & $0.555(16)$\\
		$q^2_{\mathrm{max}}$ & $0.817(11)$ & $0.911(18)$ & $0.756(16)$ & $0.910(28)$ \\
		\hline\hline
	\end{tabular}
	\label{tab:ff_extrema}
\end{table}
%

\subsection{Dependence of the form factors on the spectator quark mass}
\label{sec:ff_spectator}

\begin{figure}
\centering
\includegraphics[width=0.45\textwidth]{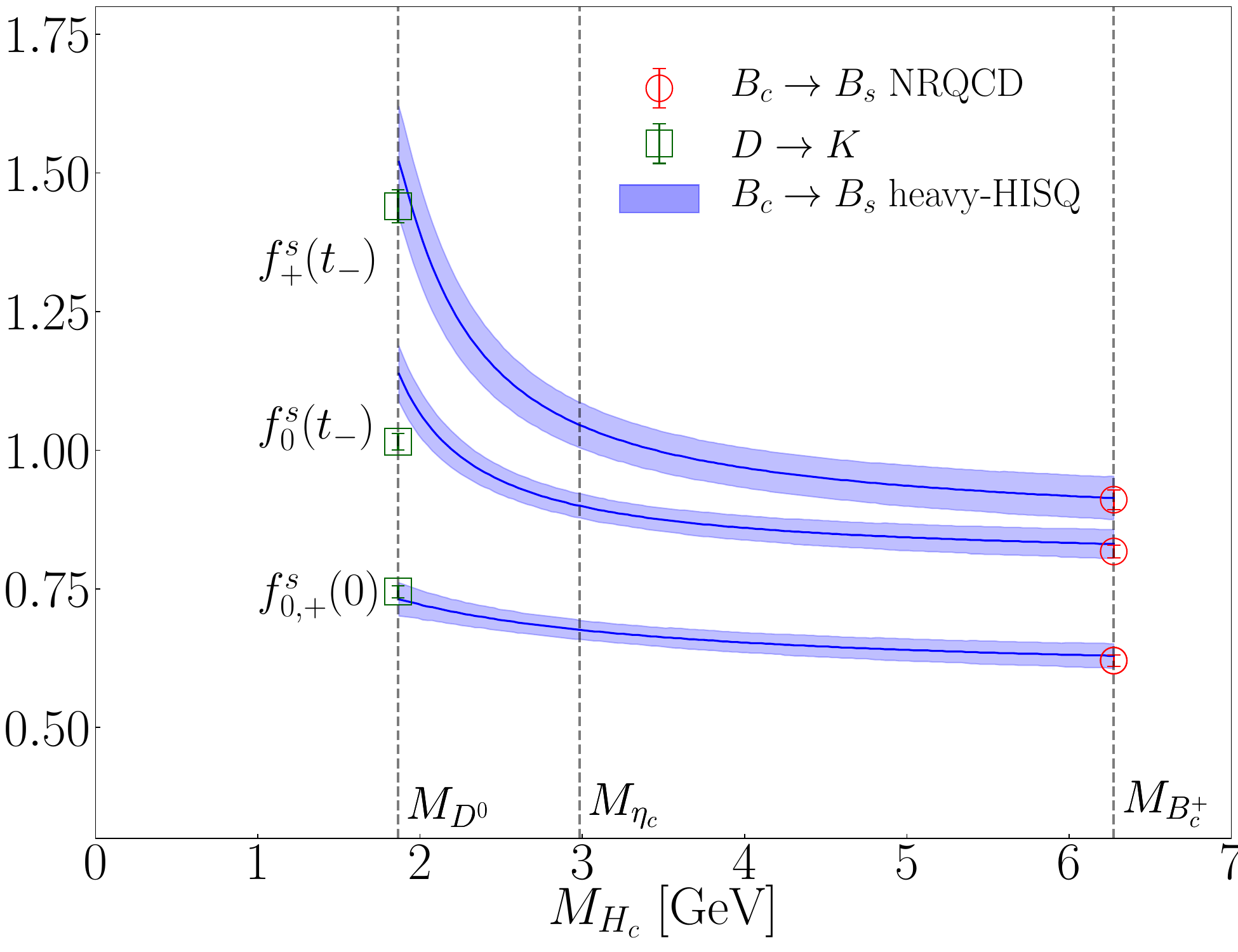}
\caption[]{Values for the physical-continuum form factors $f_0^s=f_+^s$ at $q^2=0$ and $f_0^s$ and $f_+^s$ at $q^2_\text{max}$ are plotted against the mass of the heavy-charm pseudoscalar meson. 
The curve is the continuum limit of the heavy-HISQ fit function (Eq.~(\ref{hhisqffff})) 
extrapolated to the physical $B_c$ and $D$ masses. Note that the region in which the heavy-HISQ calculation 
has results is the region above $M_{\eta_c}$. See the text for a description of how the 
extrapolation down to the $D$ was done.  
Also plotted are the form factor results for $D \to K$~\cite{Koponen:2013tua} 
(green squares) as well as the NRQCD $B_c \rightarrow B_s$ result presented in this work (red circles).}
\label{ff_zr}
\end{figure}

In order to build up a picture of the behaviour of form factors it is interesting to ask: how 
do the form factors for $c$ to $s/d$ decay depend on the mass of the spectator quark? 
We can answer that question with our heavy-HISQ calculation because we have results at a range of 
spectator quark masses from $m_c$ upwards (see Fig.~\ref{hhisqmassplot}).  
Our form factor fits (Sec.~\ref{subsec:hHISQ_ff_fits}) enable us to extrapolate up to $m_b$. 
Our most accurate results are for the $c$ to $s$ decay case and we concentrate on that here. 

Fig.~\ref{ff_zr} shows the fit curve from the heavy-HISQ results for $f^s_+$ and $f_0^s$ 
as a function of the heavy-charm meson mass (as a proxy for the spectator 
quark mass). The form factor curves that are plotted are 
those for $q^2=0$ (where $f_+=f_0$) and for the zero-recoil point ($q^2_{\text{max}}$). 
At $q^2_{max}$ the daughter meson is at rest in the rest-frame of the $H_c$ meson. 
The $q^2$ value at $q^2_{\text{max}}$ falls slowly as the heavy-quark mass increases above $m_c$ 
because the 
mass difference between $H_c$ and $H_s$ mesons falls. 
Examining the region between $M_{\eta_c}$ and $M_{B_c}$ in Fig.~\ref{ff_zr} we see almost
no dependence on the spectator mass. The form factor value that shows the most dependence is 
$f_+(q^2_{max})$. This is not surprising because $f_+$ shows the biggest slope in $q^2$ 
close to $q^2_{\text{max}}$ and hence sensitivity to the value of $q^2_{\text{max}}$.  
Note that the curve from the heavy-HISQ analysis agrees with the NRQCD results at a spectator 
mass equal to that of the $b$. As discussed in the previous subsection, the form factors 
obtained from the two calculations agree across the full $q^2$ range. 

We can also investigate the behaviour of the heavy-HISQ fit function as $m_h$ is taken below $m_c$ to $m_l$ where contact is made with results for $D \to K$ from~\cite{Koponen:2013tua}. 
For the form factors at $q^2=0$, we have $P(q^2)=1$ and our fit form at Eq.~(\ref{hhisqffff}) depends only on $M_{H_c}$. 
This permits a straightforward extrapolation to the point $M_{H_c}=M_D$ in the continuum limit.
For the form factors at zero-recoil ($q^2_\text{max}$), constructing the extrapolation curve is complicated by requiring the dependence of $q^2_{\text{max}}$ on the mass of the spectator quark. 
This requires knowledge of $M_{H_s}$ as a function of $M_{H_c}$. 
To achieve this, we fit our values of $M_{H_s}$ taken from set 6, together with physical values from experiment~\cite{PDG} at $m_h = m_l,m_b$ (i.e. $M_K$ and $M_{B_s}$), using a simple fit form $M_{H_s} = M_{H_c}(1 + \sum_{n=1}^4 \omega_n(\Lambda_{QCD}/M_{H_c})^n + A(a\Lambda_{QCD})^2+B(a\Lambda_{QCD})^4)$. 
Here $A$, $B$ and $\omega_n$ take prior values $0(2)$ and we do not include $a\Lambda$ terms for data from~\cite{PDG}.
We find this fit function reproduces our data, as well as the physical values, well. 
Fig.~\ref{ff_zr} also shows the result of this downward extrapolation.
Whilst this extrapolation below $m_c$ is outside the region where HQET is expected to be valid, the curves nevertheless show approximately the correct amount 
of upward movement necessary to reproduce the $D \rightarrow K$ 
results in~\cite{Koponen:2013tua} for $f_+$ and $f_0$ 
at zero-recoil and $q^2 = 0$.
The form factors at $q^2=0$ continue to show almost no spectator mass dependence, and this is in agreement with the $D \rightarrow K$ results.

\subsection{Decay rate}
The hadronic quantity required for determining the decay rate and branching fraction is the integral
\begin{align}
	\Gamma |V|^{-2} = \frac{G_F^2}{24 \pi^3}  \int_0^{t_-} d q^2 \hspace{1mm} |\mathbf{p}_2|^3 |f_+ (q^2)|^2,
\end{align}
where $V$ is the CKM element $V_{cs}$ or $V_{cd}$.
Table \ref{tab:GammaV} gives values for this quantity for each of the $B_c \to B_{s}$ and $B_c \to B_{d}$ processes based on the NRQCD and heavy-HISQ chained form factor fit described in Sec.~\ref{sec:ff_chained_fit}.
\begin{table}
	\centering
	\caption{Final results of the weighted integral of $|f_+ (q^2)|^2$ over the physical range of squared 4-momentum transfer. Units are $\text{MeV}$.}
	\begin{tabular}{c c c} 
		\hline\hline
		& $B_c^+ \to B_s^0 \overline{\ell} \nu_{\ell}$ & $B_c^+ \to B^0 \overline{\ell} \nu_{\ell}$ \\ [0.1ex] 
		\hline
		$\Gamma |V|^{-2}$ \hspace{5mm}& $1.738(55)\times 10^{-11}$ & $2.29(12)\times 10^{-11}$  \\
		\hline\hline
	\end{tabular}
	\label{tab:GammaV}
\end{table}
Values for different $q^2$ bins can also be obtained.
Proceeding with the total decay rate, combining these results with existing CKM matrix values \cite{PDG} $V_{cs} = 0.997(17)$ and $V_{cd} = 0.218(4)$ yields the predictions
\begin{align}
	\Gamma (B_c^+ \to B_s^0 \overline{\ell} \nu_{\ell}) 
	&= 26.25(90)(83) \times 10^{9} \hspace{1mm}\text{s}^{-1} \nonumber \\
	\Gamma (B_c^+ \to B^0 \overline{\ell} \nu_{\ell})
	&= 1.650(61)(84)\times 10^{9}\hspace{1mm} \text{s}^{-1}
\end{align}
where the CKM matrix elements are responsible for the first errors and the second errors arise from our lattice calculations.
The dominant source of lattice QCD uncertainty is the fitting of 2-point and 3-point correlators described in Sec.~\ref{2ptcorrels}.

We can convert these results for the decay width into a branching fraction using 
the lifetime of the $B_c$ meson, $513.49(12.4) \; \mathrm{fs}$~\cite{Aaij:2014gka}.  
This gives 
\begin{align}
	\mathcal{B} (B_c^+ \to B_s^0 \overline{\ell} \nu_{\ell}) 
	&= 0.01348(46)(33)(43) \nonumber \\
	\mathcal{B} (B_c^+ \to B^0 \overline{\ell} \nu_{\ell})
	&= 0.000847(31)(43)(20)
\end{align}
where now the third uncertainty is from the lifetime. 

We also present the ratio of the $\Gamma|V|^{-2}$ for $B_c \rightarrow B_s$ to $B_c \rightarrow B_d$ 
taking correlations into account between the numerator and denominator. 
From the chained fit of $B_c^+ \to B_s^0 \overline{\ell} \nu_{\ell}$ and $ B_c^+ \to B^0 \overline{\ell} \nu_{\ell}$ form factors, we obtain
\begin{align}
	\frac{\Gamma (B_c^+ \to B_s^0 \overline{\ell} \nu_{\ell}) |V_{cd}|^2}{ \Gamma (B_c^+ \to B^0 \overline{\ell} \nu_{\ell}) |V_{cs}|^2}  = 0.759(44).
\end{align}
In fact the uncertainty is roughly the same as if we were to treat the numerator and denominator as uncorrelated.

\begin{figure}
	\centering
	\includegraphics[width=0.45\textwidth]{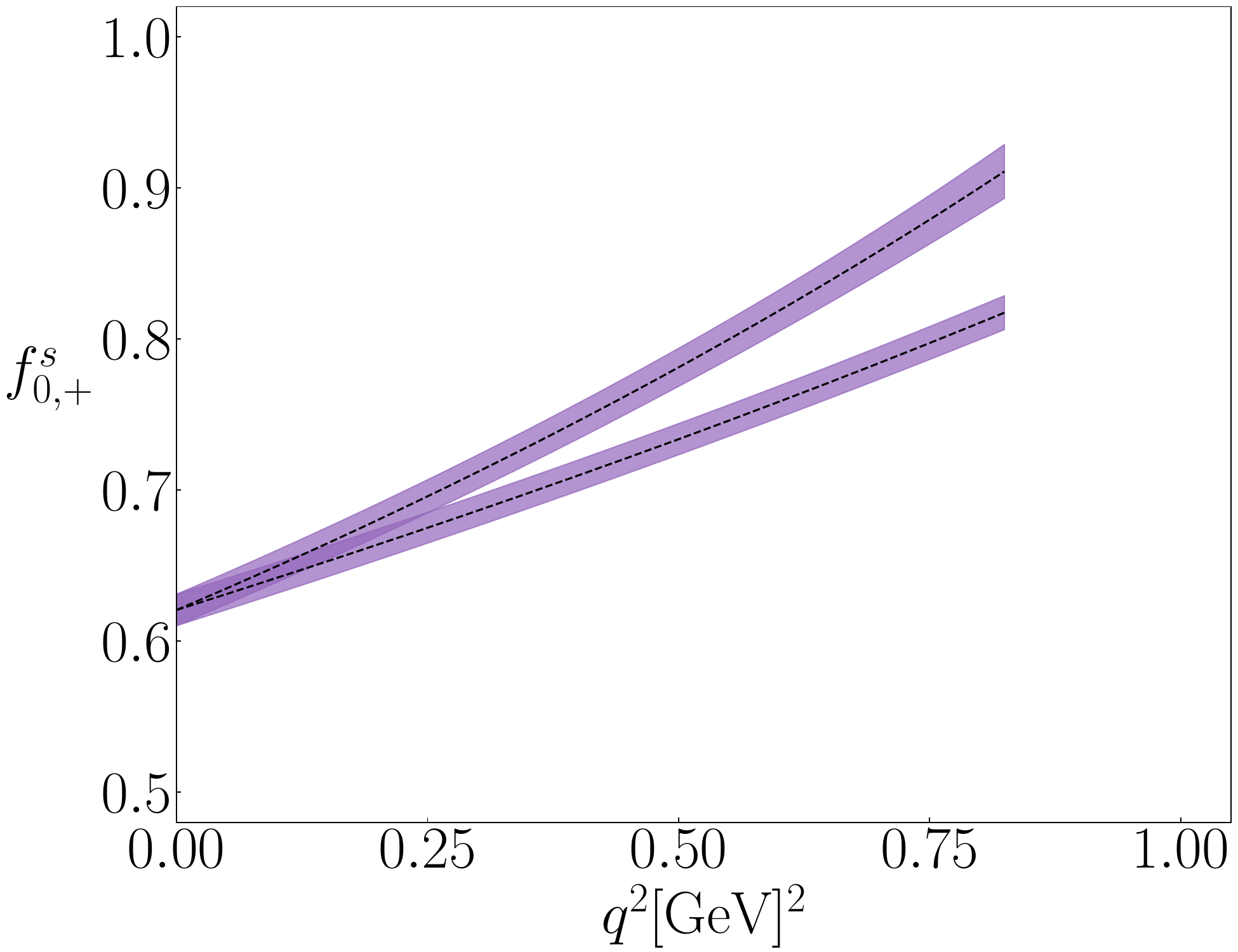}
	\caption{Final form factors from the chained fits of $f_0$ (below) and $f_+$ (above) for $B_c^+ \to B_s^0 \overline{\ell} \nu_{\ell}$ 
		in the physical-continuum limit, plotted against the entire range of physical $q^2$. This fit is described in Sec.~\ref{sec:ff_chained_fit}.}
	\label{ff_strange_chainedOnly}
\end{figure}
\begin{figure}
	\centering
	\includegraphics[width=0.45\textwidth]{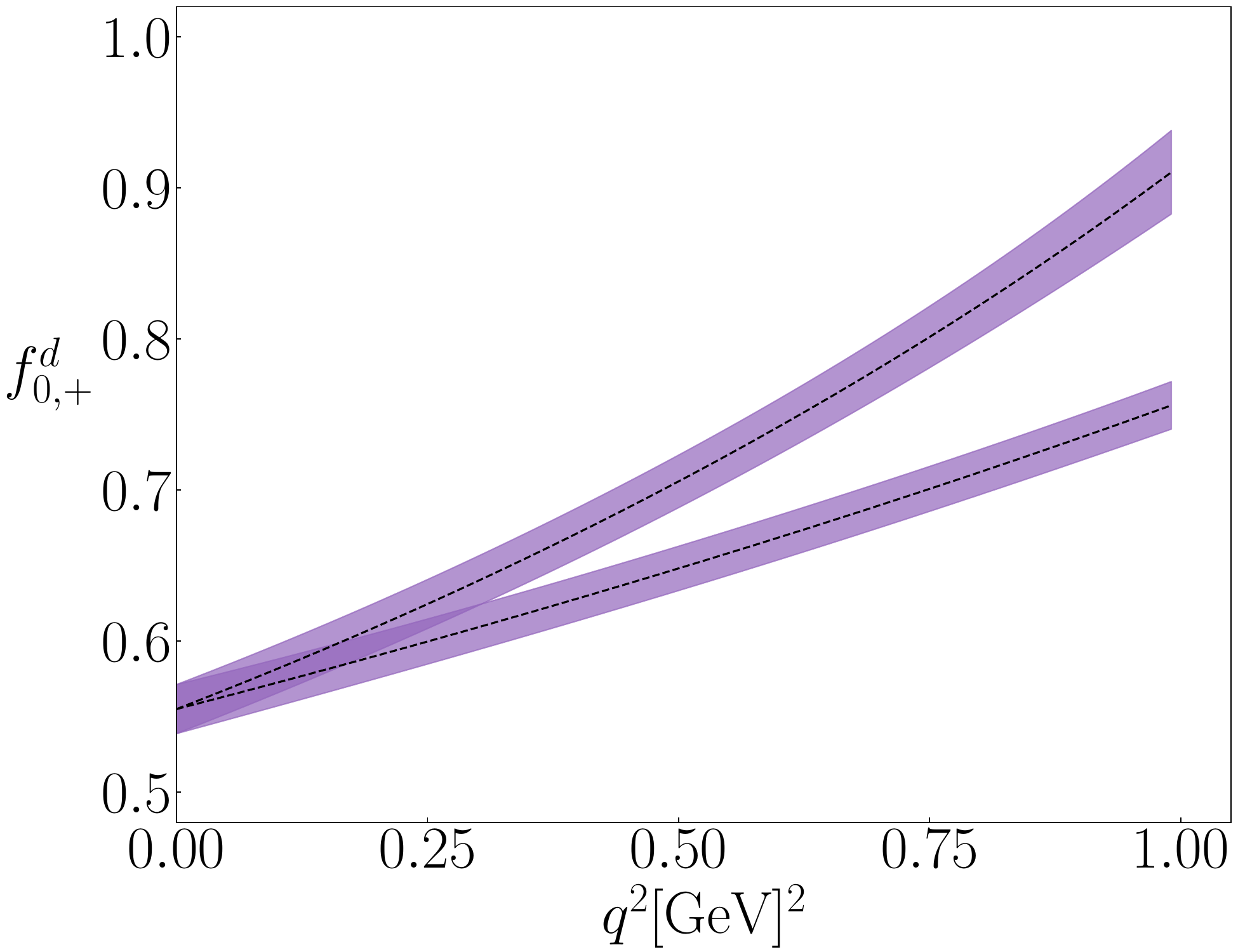}
	\caption{Final form factors from the chained fits of $f_0$ (below) and $f_+$ (above) for $B_c^+ \to B^0 \overline{\ell} \nu_{\ell}$ in the physical-continuum limit, plotted against the entire range of physical $q^2$. 
		This fit is described in Sec.~\ref{sec:ff_chained_fit}.}
	\label{ff_down_chainedOnly}
\end{figure}
\section{Conclusions}
\label{sec:conclusions}

We have reported here the first calculations of the decay rates $\Gamma (B_c^+ \to B_s^0 \overline{\ell} \nu_{\ell})$ and $\Gamma (B_c^+ \to B^0 \overline{\ell} \nu_{\ell})$, demonstrating the success of lattice QCD in studying decays of heavy-light mesons.
The use of HISQ-HISQ $c \to s(d)$ currents allows for a non-perturbative renormalisation using the PCVC.
We used two different formulations for the spectator $b$ quark, heavy-HISQ and NRQCD. 
Results from the heavy-HISQ calculations are in good agreement with the 
physical-continuum form factors derived from the calculations using NRQCD $b$ quarks, 
giving us confidence in assessing and controlling the systematic errors in each formulation.
Simulating at a variety of spectator masses in the heavy-HISQ calculation has 
provided a check of the spectator-independence of the renormalisation procedure for the vector 
current. The NRQCD study also accessed $Z_V$ away from zero-recoil to scrutinise momentum independence.

Our final form factors from the chained fit that combines both NRQCD and heavy-HISQ 
results are plotted against $q^2$ in Figs.~\ref{ff_strange_chainedOnly} and~\ref{ff_down_chainedOnly}.

The decay rates are predicted from our calculation with 4.6\% and 6.3\% uncertainty 
for $\Gamma (B_c^+ \to B_s^0 \overline{\ell} \nu_{\ell})=26.2(1.2) \times 10^{9} \hspace{1mm}\text{s}^{-1}$ and $\Gamma (B_c^+ \to B^0 \overline{\ell} \nu_{\ell})=1.65(10)\times 10^{9}\hspace{1mm} \text{s}^{-1}$ respectively. 
There is scope for significant improvement should future experiment demand more precision from the lattice.
Such improvement would be readily achieved by the inclusion of lattices with a finer lattice 
in the heavy-HISQ calculation. `Ultrafine' lattices with $a \approx 0.045$ [fm] were used 
in~\cite{McLean:2019qcx} to provide results nearer to the physical-continuum limit 
with $am_h \approx am_b$. Larger statistical samples could also be obtained on the lattices used 
here, at the cost of more computational resources.


\section*{ACKNOWLEDGMENTS}

We are grateful to Mika Vesterinen for asking us about the form factors for these decays at the UK Flavour 2017 workshop at the IPPP, Durham.
We are also grateful to Matthew Kenzie for discussions about the prospects of measurements by LHCb.
We thank Jonna Koponen, Andrew Lytle and Andre Zimermmane-Santos for making previously generated lattice propagators available for our use and Euan McLean for useful discussions on setting up the calculations.
We thank the MILC Collaboration for making publicly available their gauge configurations and their code MILC-7.7.11 \cite{MILCgithub}.
This work was performed using the Cambridge Service for Data Driven Discovery (CSD3), part of which is operated by the University of Cambridge Research Computing on behalf of the STFC DiRAC HPC Facility (www.dirac.ac.uk).
The DiRAC component of CSD3 was funded by BEIS capital funding via STFC capital grants ST/P002307/1 and ST/R002452/1 and STFC operations grant ST/R00689X/1.
DiRAC is part of the National e-Infrastructure.
We are grateful to the CSD3 support staff for assistance.
This work has been supported by STFC consolidated grants ST/P000681/1 and ST/P000746/1.

\bibliography{BcBs_bib}{}
\bibliographystyle{apsrev4-1}

\end{document}